\documentstyle[12pt,fleqn,epsf,epsfig,axodraw]{article}
\renewcommand{\theequation}{\arabic{equation}}

\def\lsim{\raise0.3ex\hbox{$\;<$\kern-0.75em\raise-1.1ex\hbox{$\sim\;$}}}
\def\gsim{\raise0.3ex\hbox{$\;>$\kern-0.75em\raise-1.1ex\hbox{$\sim\;$}}}
\newcommand{\slashed}[1]{\not\!#1}

\begin{document}
\setlength{\unitlength}{1cm}
\setlength{\mathindent}{0cm}
\thispagestyle{empty}
\null
\hfill WUE-ITP-04-005\\
\null
\hfill IFIC-04-02\\
\null
\hfill UWThPh-2004-01\\
\null
\hfill HEPHY-PUB 783/04\\
\null
\hfill hep-ph/0402016\\
\vskip .8cm
\begin{center}
	{\Large \bf 
		CP sensitive observables in  
		$e^+e^-\to\tilde\chi^0_i \tilde\chi^0_j$
		and neutralino decay into the $Z$ boson}
\vskip 2.5em
{\large
{\sc A.~Bartl$^{a}$\footnote{e-mail:
        bartl@ap.univie.ac.at}
     H.~Fraas$^{b}$\footnote{e-mail:
        fraas@physik.uni-wuerzburg.de},
	  O.~Kittel$^{b,c}$\footnote{e-mail:
		  kittel@physik.uni-wuerzburg.de},
	  W.~Majerotto$^{d}$\footnote{e-mail:
        majer@qhepu3.oeaw.ac.at}
}}\\[1ex]
{\normalsize \it
$^{a}$ Institut f\"ur Theoretische Physik, Universit\"at Wien, 
Boltzmanngasse 5, A-1090 Wien, Austria}\\
{\normalsize \it
$^{b}$ Institut f\"ur Theoretische Physik, Universit\"at
W\"urzburg, Am Hubland, D-97074~W\"urzburg, Germany}\\
{\normalsize \it
$^{c}$ Institut de F\'{\i}sica Corpuscular - C.S.I.C., 
Universitat de Val{\`e}ncia 
Edifici Instituts d'Investigaci{\'o}, 
- Apartat de Correus 22085 - 
E-46071 Val{\`e}ncia, Spain}\\
{\normalsize \it
$^{d}$ Institut f\"ur Hochenergiephysik, \"Osterreichische
Akademie der Wissenschaften, Nikolsdorfergasse 18, 
A-1050 Wien, Austria}\\
\vskip 1em
\end{center} \par
\vskip .8cm

\begin{abstract}

We study CP sensitive observables in neutralino production 
$e^+e^- \to\tilde\chi^0_i \tilde\chi^0_j$
and the subsequent two-body decays of the neutralino
$\tilde\chi^0_i \to \chi^0_n Z$ and of the $Z$ boson
$Z \to \ell \bar\ell (q\bar q)$.
We identify the CP odd  elements of the $Z$ boson
density matrix and  propose CP sensitive 
triple-product asymmetries. 
We calculate these observables and the cross sections 
in the Minimal Supersymmetric Standard Model with complex parameters 
$\mu$ and $M_1$ for  an $e^+e^-$ linear collider with 
$\sqrt{s}=800$ GeV and longitudinally polarized beams. 
We show that the asymmetries can reach  $3\%$ 
for $Z \to \ell \bar\ell $ and
$18\%$ for $Z \to  q\bar q$
and discuss the feasibility of measuring these asymmetries.

\end{abstract}

\newpage

\section{Introduction}

In the Minimal Supersymmetric Standard Model (MSSM) \cite{haberkane} 
several supersymmetric (SUSY) parameters  can be complex. 
In the neutralino sector of the MSSM these are the $U(1)$ gaugino mass 
parameter $M_1$ and the Higgsino mass parameter $\mu$. 
(The $SU(2)$ gaugino mass parameter $M_2$  can be made real 
by redefining the fields.) The physical phases $\varphi_{M_1}$
and $\varphi_{\mu}$ of $M_1$ and $\mu$, respectively, 
imply CP odd observables which can in principle be large, 
because they are already present at tree level. 
It has been shown  that in the production  of two different neutralinos
$e^+e^-\to\tilde\chi^0_i\tilde\chi^0_j$ the CP violating phases
cause a non-vanishing neutralino polarization 
perpendicular to the production plane \cite{kali,gudi1,oshimo,choi1},
which leads to CP odd triple-product asymmetries \cite{tripleprods}
of the neutralino decay products
\cite{oshimo,choi1,olaf1,drees,staudecay}.

In this work we study CP violation in neutralino production
\begin{eqnarray} \label{productionA}
	e^++e^-&\to&\tilde\chi^0_i+\tilde\chi^0_j, 
	\quad i,j =1,\dots,4,  
\end{eqnarray}
with the subsequent  two-body decay of one neutralino into the $Z$ boson
(for recent studies see \cite{staudecay,choi2})
\begin{eqnarray} \label{decay_1A}
     \tilde\chi^0_i \to \chi^0_n +Z; \quad n<i,
\end{eqnarray}
and the decay of the $Z$ boson
	\begin{eqnarray} \label{decay_2B}
		Z \to f +\bar f , \quad  f=\ell,q,\quad\ell=e,\mu,\tau, \quad q=c,b.
\end{eqnarray}
In case of  CP violation the non-vanishing phases 
$\varphi_{M_1}$ and $\varphi_{\mu}$ lead to CP sensitive elements
of the $Z$ boson density matrix, which we will discuss
in detail. Moreover, these CP sensitive elements cause
CP odd asymmetries ${\mathcal A}_{f}$ in the decay 
distribution of the decay fermions \cite{oshimo}:
 \begin{eqnarray}\label{AT1}
	 {\mathcal A}_{f} &=& 
	 \frac{\sigma({\mathcal T}_{f}>0)-\sigma({\mathcal T}_{f}<0)}
	{\sigma({\mathcal T}_{f}>0)+\sigma({\mathcal T}_{f}<0)},
\end{eqnarray}
with $\sigma$ the cross section and the triple product
 \begin{eqnarray}\label{tripleproduct}
	 {\mathcal T}_{f} &=& 
	 \vec p_{e^-}\cdot(\vec p_{f} \times \vec p_{\bar f}).
 \end{eqnarray}
Due to the correlations between the $\tilde\chi^0_i$ 
polarization and the $Z$ boson polarization, there are 
CP odd contributions to the $Z$ boson density matrix
and to the asymmetries from the production (\ref{productionA}) 
and  from the decay process  (\ref{decay_1A}).
 
The triple product ${\mathcal T}_{f}$, Eq. (\ref{tripleproduct}) 
changes sign under time reversal and is thus T odd. Due to CPT invariance, 
the corresponding T odd asymmetries ${\mathcal A}_{f}$ 
are also CP odd if final state
interactions are neglected.
The final state interactions would also contribute to 
${\mathcal A}_{f}$. However, they only arise at loop level and 
are neglected in the present work.

In Section \ref{Definitions and Formalism} we give our definitions and 
the formalism used and define the $Z$ boson density matrix. 
In Section \ref{T odd asymmetry} we discuss some general
properties of the asymmetries. We present numerical results  
in Section \ref{Numerical results}. 
Section \ref{Summary and conclusion} gives a summary 
and conclusions.

\section{Definitions and formalism
  \label{Definitions and Formalism}}

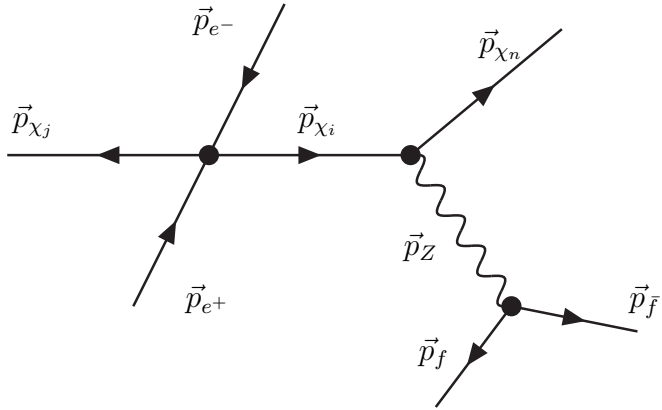
\begin{figure}[h]
\begin{picture}(5,6.)(-2,.5)
		\put(1,4.7){$\vec p_{\chi_j}$}
   \put(3.4,6){$\vec p_{e^- }$}
   \put(3.3,2.3){$\vec  p_{e^+}$}
   \put(4.8,4.7){$\vec p_{\chi_i}$}
   \put(7.2,5.7){$ \vec p_{\chi_n}$}
   \put(6.2,3.){$ \vec p_{Z}$}
   \put(9.2,2.3){$\vec  p_{\bar f}$}
   \put(6.4,1.6){$ \vec p_{f}$}
\end{picture}
\scalebox{1.9}{
\begin{picture}(0,0)(1.3,-0.25)
\ArrowLine(40,50)(0,50)
\Vertex(40,50){2}
\ArrowLine(55,80)(40,50)
\ArrowLine(25,20)(40,50)
\ArrowLine(40,50)(80,50)
\ArrowLine(80,50)(110,75)
\Photon(80,50)(100,20){2}{5}
\Vertex(80,50){2}
\ArrowLine(100,20)(125,15)
\ArrowLine(100,20)(85,0)
\Vertex(100,20){2}
\end{picture}}
\caption{\label{shematic picture}
          Schematic picture of the neutralino production
          and decay process.}
\end{figure}

We give the analytic formulae for the differential cross section
of neutralino production 
\begin{eqnarray} \label{production}
	e^++e^-&\to&\tilde\chi^0_i(p_{\chi_i}, \lambda_i)+
	            \tilde\chi^0_j(p_{\chi_j}, \lambda_j), 
\end{eqnarray}
with longitudinally polarized beams and the subsequent decay chain
of one of the neutralinos
\begin{eqnarray} \label{decay_1}
	\tilde\chi^0_i &\to&\tilde \chi^0_n(p_{\chi_n}, \lambda_n) 
	    +Z(p_Z, \lambda_k), \\
	  Z &\to& f(p_{f}, \lambda_f) +\bar f(p_{\bar f}, \lambda_{\bar f}). 
\label{decay_2}
\end{eqnarray}
In Eq.~(\ref{production}) and Eq.~(\ref{decay_1}),(\ref{decay_2}), 
$p$ and $\lambda$ denote momentum and helicity, respectively.
For a schematic picture of the neutralino production and decay process 
see Fig.~\ref{shematic picture}.
In the following we will derive the 
$Z$ boson spin-density matrix and relate it to
the CP asymmetry ${\mathcal A}_{f}$ in Eq.~(\ref{AT1}).

\subsection{Lagrangian and helicity amplitudes
     \label{Lagrangian}}

The interaction Lagrangians relevant for our study are 
(in our notation and conventions we follow 
closely \cite{haberkane,staudecay}):
\begin{eqnarray}
{\cal L}_{Z^0\tilde{\chi}^0_i\tilde{\chi}^0_j} &=&
{\textstyle\frac{1}{2}}Z_{\mu}\bar{\tilde{\chi}}^0_i\gamma^{\mu}
[O_{ij}^{''L} P_L+O_{ij}^{''R} P_R]\tilde{\chi}^0_j, \quad i, j=1,\dots,4, \\
{\cal L}_{e \tilde{e}\tilde{\chi}^0_i} &=&
g f^L_{e i}\bar{e}P_R\tilde{\chi}^0_i\tilde{e}_L+
g f^R_{e i}\bar{e}P_L\tilde{\chi}^0_i\tilde{e}_R+\mbox{h.c.}, \\
{\cal L}_{Z^0 f \bar f} &=&
Z_{\mu}\bar f\gamma^{\mu}[L_fP_L+ R_f P_R]f,
\end{eqnarray}
with $P_{L, R}=\frac{1}{2}(1\mp \gamma_5)$.
In the neutralino basis $\tilde{\gamma},
\tilde{Z}, \tilde{H}^0_a, \tilde{H}^0_b$ the couplings are:
\begin{eqnarray}
	&&O_{ij}^{''L}=-\frac{1}{2}\frac{g}{\cos\theta_W} \left[
	(N_{i3}N_{j3}^*-N_{i4}N_{j4}^*)\cos2\beta
  +(N_{i3}N_{j4}^*+N_{i4}N_{j3}^*)\sin2\beta \right],\\
&&O_{ij}^{''R}=-O_{ij}^{''L*},\quad\quad
L_f=-\frac{g}{\cos\theta_W}(T_{3f}-q_{f}\sin^2\theta_W), \quad
R_f=\frac{g}{\cos\theta_W}q_{f}\sin^2\theta_W \label{eq_5},\\
&&f_{\ell i}^L = -\sqrt{2}\bigg[\frac{1}{\cos
\theta_W}(T_{3\ell}-q_{\ell}\sin^2\theta_W)N_{i2}+q_{\ell}\sin \theta_W
N_{i1}\bigg],\\
&&f_{\ell i}^R = -\sqrt{2}q_{\ell} \sin \theta_W
\Big[\tan \theta_W N_{i2}^*- N_{i1}^*\Big],
\label{eq_6}
\end{eqnarray}
with $g$  the weak coupling constant ($g=e/\sin\theta_W$, $e>0$),
$q_f$ and $T_{3f}$  the 
charge and the isospin of the fermion, and
$\tan \beta=v_2/v_1$ the ratio of the vacuum expectation values 
of the two neutral Higgs fields.
$N_{ij}$ is the complex unitary $4\times 4$ matrix which diagonalizes
the neutral gaugino-Higgsino mass matrix $Y_{\alpha\beta}$, 
 $N_{i \alpha}^*Y_{\alpha\beta}N_{\beta k}^{\dagger}=
 m_{\tilde{\chi}^0_i}\delta_{ik}$, with $ m_{\tilde{\chi}^0_i}>0$.
Note that our definitions of $O_{ij}^{''L,R}$ and $L_f,R_f$ differ
from those given in \cite{haberkane,gudi1} by a factor of $g/\cos\theta_W$.

The helicity amplitudes
$T_P^{\lambda_i \lambda_j}$
for the production process are given in \cite{gudi1}.
Those for the two-body decays, 
Eq.~(\ref{decay_1}) and Eq.~(\ref{decay_2}), are
\begin{eqnarray}
	T_{D_1,\lambda_i}^{\lambda_n\lambda_k} &=& 
	\bar u(p_{\chi_n},\lambda_n)
	\gamma^{\mu}[O_{ni}^{''L}P_L + O_{ni}^{''R}P_R]
		u(p_{\chi_i}, \lambda_i)
	\varepsilon_{\mu}^{\lambda_k\ast}
\end{eqnarray}
and
\begin{eqnarray}
	T_{D_2,\lambda_k}^{\lambda_f \lambda_{\bar f}} &=& 
	\bar u(p_f,\lambda_f)
	\gamma^{\mu}[L_f P_L + R_f P_R] v(p_{\bar f},\lambda_{\bar f})
	\varepsilon_{\mu}^{\lambda_k}.
\end{eqnarray}
The polarization vectors 
$\varepsilon_{\mu}^{\lambda_k},\lambda_k =0,\pm1$, are given in 
Appendix~\ref{Representation of momentum and spin vectors}.
The amplitude for the whole process 
(\ref{production}), (\ref{decay_1}), (\ref{decay_2}) is
\begin{eqnarray}
T &=& \Delta(\tilde\chi^0_i) \Delta(Z)
\sum_{\lambda_i, \lambda_k}
T_P^{\lambda_i \lambda_j}T_{D_1,\lambda_i}^{\lambda_n\lambda_k}
T_{D_2,\lambda_k}^{\lambda_f \lambda_{\bar f}},
\end{eqnarray}
with the neutralino propagator 
$ \Delta(\tilde{\chi}^0_i)=i/[p_{\chi_i}^2-m_{\chi_i}^2
	+im_{\chi_i}\Gamma_{\chi_i}]$ 
and the $Z$ boson propagator 
$ \Delta(Z)=i/[p_Z^2-m_Z^2 +im_Z \Gamma_Z]$
(the  mass and width  are denoted by  $m$ and $\Gamma$, respectively).
For these propagators we use the narrow width approximation.

\subsection{Cross section and $Z$ boson density matrix
     \label{Amplitude squared}}

For the calculation of the cross section for the
combined process of neutralino production (\ref{production})
and the subsequent two-body decays
(\ref{decay_1}), (\ref{decay_2}) of $\tilde{\chi}^0_i$
we use the same spin-density matrix formalism as in \cite{gudi1,spin}.
The (unnormalized)  spin-density matrix of the $Z$ boson 
\begin{eqnarray}       \label{Zdensitymatrix}
\rho_{P}(Z)^{\lambda_k\lambda'_k}&=&
|\Delta(\tilde\chi^0_i)|^2~
\sum_{\lambda_i,\lambda'_i}~
\rho_P   (\tilde\chi^0_i)^{\lambda_i \lambda_i'}\;
\rho_{D1}(\tilde\chi^0_i)_{\lambda_i'\lambda_i}^{\lambda_k\lambda'_k},
\end{eqnarray}
is composed of the spin-density production matrix
\begin{eqnarray} 
	\rho_P(\tilde{\chi}^{0}_i)^{\lambda_i \lambda_i'}&=&\sum_{\lambda_j}
	T_P^{\lambda_i \lambda_j}T_P^{\lambda_i' \lambda_j \ast}
\end{eqnarray}
and the decay matrix
\begin{eqnarray}
\rho_{D1}(\tilde\chi^0_i)_{\lambda_i' \lambda_i}^{\lambda_k\lambda'_k} &=&
\sum_{\lambda_n}T_{D_1,\lambda_i}^{\lambda_n\lambda_k}
T_{D_1,\lambda_i'}^{\lambda_n\lambda_k'\ast}.
\end{eqnarray}
With the decay matrix for the $Z$ decay
\begin{eqnarray}
	\rho_{D2}(Z)_{\lambda_k' \lambda_k}&=&
	\sum_{\lambda_f, \lambda_{\bar f}}
	T_{D_2,\lambda_k}^{\lambda_f \lambda_{\bar f} }
	T_{D_2,\lambda_k'}^{\lambda_f \lambda_{\bar f} \ast}
\end{eqnarray}
the amplitude squared for the complete process
$ e^+e^-\to\tilde\chi^0_i\tilde\chi^0_j$;
$\tilde\chi^0_i\to\tilde\chi^0_n Z $;
$Z \to f \bar f$ 
can now be written
\begin{eqnarray}       \label{amplitude}
|T|^2&=&|\Delta(Z)|^2
	\sum_{\lambda_k,\lambda'_k}~
	\rho_{P}(Z)^{\lambda_k\lambda'_k}\;
	\rho_{D2}(Z)_{\lambda'_k\lambda_k}.
\end{eqnarray}
The differential cross section 
in the laboratory system is then given by 
\begin{equation}\label{crossection}
	d\sigma=\frac{1}{2 s}|T|^2 
	d{\rm Lips}(s,p_{\chi_j },p_{\chi_n},p_{f},p_{\bar f}),
\end{equation}
where $d{\rm Lips}(s,p_{\chi_j },p_{\chi_n},p_{f},p_{\bar f})$
is the Lorentz invariant phase space element
defined in  Eq.~(\ref{Lips}) of Appendix \ref{Phase space}.
More details concerning kinematics and phase space  
can be found in Appendices \ref{Representation of momentum and spin vectors}
and \ref{Phase space}.

For the polarization of the decaying neutralino $ \tilde \chi^0_i$
with momentum $p_{\chi_i}$ we introduce 
three space like spin vectors
$s^a_{\chi_i}\;(a=1,2,3)$, 
which together with $p_{\chi_i}^{\mu}/m_{\chi_i}$
form an orthonormal set  with 
$s^a_{\chi_i}\cdot s^b_{\chi_i}=-\delta^{ab}$, 
$s^a_{\chi_i}\cdot p_{\chi_i}=0$,
then the  (unnormalized) neutralino density matrix can be expanded 
in terms of the Pauli matrices:
\begin{eqnarray} \label{rhoP}
  \rho_P(\tilde\chi^0_i)^{\lambda_i \lambda_i'} &=&
      2(\delta_{\lambda_i \lambda_i'} P + 
       \sigma^{a}_{\lambda_i \lambda_i'}
		\Sigma_P^a),   
\end{eqnarray}
where we sum over $a$.
With our choice of the spin vectors $s^a_{\chi_i}$,
given in Appendix~\ref{Representation of momentum and spin vectors},
$\frac{\Sigma^{3}_P}{P}$
is the longitudinal polarization of neutralino $ \tilde \chi^0_i$,
$\frac{\Sigma^{1}_P}{P}$ is the transverse polarization in the 
production plane and $\frac{\Sigma^{2}_P}{P}$ is the polarization
perpendicular to the production plane.
The  analytical formulae for $P$ and $\Sigma^{a}_P$ are given in \cite{gudi1}.
To describe the polarization states of the $Z$ boson,
we introduce  a set of spin vectors $t^c_Z\;(c=1,2,3)$ and choose
polarization vectors $\varepsilon^{\lambda_k}_{\mu} (\lambda_k=0,\pm1)$, 
given in Appendix~\ref{Representation of momentum and spin vectors}.
Then we obtain for the decay matrices 
\begin{eqnarray} \label{rhoD1}
\rho_{D1}(\tilde\chi^0_i)_{\lambda_i' \lambda_i}^{\lambda_k\lambda'_k} &=& 
(\delta_{\lambda_i' \lambda_i} D_1^{\mu\nu} 
+ \sigma^a_{\lambda_i'\lambda_i}  \Sigma^{a~\mu\nu}_{D1})
		  \varepsilon_{\mu}^{\lambda_k\ast}\varepsilon_{\nu}^{\lambda'_k}
\end{eqnarray}
and
\begin{eqnarray}\label{rhoD2}
		  \rho_{D2}(Z)_{\lambda'_k\lambda_k}&=& D_2^{\mu\nu}
\varepsilon_{\mu}^{\lambda_k}\varepsilon_{\nu}^{\lambda'_k\ast},
\end{eqnarray}
with \cite{staudecay}:
\begin{eqnarray}
	D_1^{\mu\nu} &=&
	2[2 p^{\mu}_{\chi_i} p^{\nu}_{\chi_i}
		-(p^{\mu}_{\chi_i}p^{\nu}_Z +  p^{\nu}_{\chi_i} p^{\mu}_Z)
		- {\textstyle\frac{1}{2}}(m_{\chi_i}^2+m_{\chi_n}^2-m_Z^2)g^{\mu\nu} ]
	|O^{''L}_{ni}|^2 \nonumber\\
	&&-2g^{\mu\nu}m_{\chi_i}m_{\chi_n}[(Re O^{''L}_{ni})^2 -(Im
			O^{''L}_{ni})^2], \\
\Sigma_{D_1}^{a~ \mu\nu} &=& 
	2i\{ - m_{\chi_i}\epsilon^{\mu\alpha\nu\beta} 
	s^a_{\chi_i\alpha}(p_{\chi_i\beta} -p_{Z\beta} )|O^{''L}_{ni}|^2
	+2m_{\chi_n}(s^{a\mu}_{\chi_i}p^{\nu}_{\chi_i}-s^{a\nu}_{\chi_i}p^{\mu}_{\chi_i})
	(Im O^{''L}_{ni})(Re O^{''L}_{ni})\nonumber\\
	&&-m_{\chi_n}\epsilon^{\mu\alpha\nu\beta}s^a_{\chi_i\alpha}p_{\chi_i\beta} 
	[(Re O^{''L}_{ni})^2 -(Im O^{''L}_{ni})^2]
		\}; \quad (\epsilon_{0123}=1),
\end{eqnarray}
and
\begin{eqnarray}
	D_2^{\mu\nu} &=&2(-2p^{\mu}_{\bar f}p^{\nu}_{\bar f} 
		+p^{\mu}_Zp^{\nu}_{\bar f} +p^{\mu}_{\bar f}p^{\nu}_Z
		-{\textstyle\frac{1}{2}}m_Z^2 g^{\mu\nu})(L_f^2+R_f^2)
	-2i\epsilon^{\mu\alpha\nu\beta}p_{Z\alpha}p_{\bar f\beta}(L_f^2-R_f^2).
\end{eqnarray}
Due to the Majorana properties of the neutralinos,
$D_1^{\mu\nu}$ is symmetric and $\Sigma^{a~\mu\nu}_{D1}$
is antisymmetric  under interchange of $\mu$ and $\nu$.
In Eq.~(\ref{rhoD1}) and  Eq.~(\ref{rhoD2}) we use the 
expansion \cite{choiBM}:
\begin{eqnarray}\label{expansion}
\varepsilon_{\mu}^{\lambda_k}\varepsilon_{\nu}^{\lambda'_k\ast}&=&
{\textstyle \frac{1}{3}}\delta^{\lambda_k'\lambda_k}I_{\mu\nu}
-\frac{i}{2m_Z}\epsilon_{\mu\nu\rho\sigma}
p_Z^{\rho}t_Z^{c\sigma}(J^c)^{\lambda_k'\lambda_k}
-{\textstyle \frac{1}{2}}t_{Z\mu}^c t_{Z\nu}^d (J^{cd})^{\lambda_k'\lambda_k},
\quad(\epsilon_{0123}=1),
\end{eqnarray}
summed over $c,d$. Here, $J^c$ are the $3\times3$ spin 1 matrices with
$[J^c,J^d]=i\epsilon_{cde}J^e$ and
\begin{eqnarray}
J^{cd}&=&J^cJ^d+J^dJ^c-{\textstyle \frac{4}{3}}\delta^{cd},
\end{eqnarray}
with $J^{11}+J^{22}+J^{33}=0$, are the components of a symmetric,
traceless tensor, given in Appendix~\ref{matrices}, and 
\begin{eqnarray}
I_{\mu\nu}&=&-g_{\mu\nu}+\frac{p_{Z \mu}p_{Z \nu}}{m_Z^2}
\end{eqnarray}
guarantees the completeness relation of the polarization vectors
\begin{eqnarray}\label{completeness}
\sum_{\lambda_k} \varepsilon^{\lambda_k\ast}_{\mu}
\varepsilon^{\lambda_k}_{\nu}&=& -g_{\mu\nu}+\frac{p_{Z \mu}p_{Z \nu}}{m_Z^2}.
\end{eqnarray}
The second term in Eq.~(\ref{expansion}) describes the vector
polarization and the third term describes the tensor polarization
of the spin 1 $Z$ boson.
The decay matrices can be expanded in terms of the spin matrices
$J^c$ and $J^{cd}$. The first term  of the decay
matrix $\rho_{D1}$, Eq.~(\ref{rhoD1}), 
which is independent of the neutralino polarization, then gives
\begin{eqnarray}
D_1^{\mu\nu} 
\varepsilon_{\mu}^{\lambda_k\ast}\varepsilon_{\nu}^{\lambda'_k}&=&
  D_1 \delta^{\lambda_k\lambda_k'} 
  + \,^cD_1(J^c)^{\lambda_k\lambda_k'}
  + \,^{cd}D_1(J^{cd})^{\lambda_k\lambda_k'},
\end{eqnarray}
summed over $c,d$, with
\begin{eqnarray}\label{D1}
D_1&=&\Big[m_{\chi_n}^2-{\textstyle \frac{1}{3}}m_{\chi_i}^2-m_Z^2+\frac{4}{3}
		\frac{(p_{\chi_i}\cdot p_Z)^2}{m_Z^2}\Big]|O^{''L}_{ni}|^2\nonumber\\
&&+2m_{\chi_i}m_{\chi_n}[(Re O^{''L}_{ni})^2 -(Im
			O^{''L}_{ni})^2],\\
^{cd}D_1&=&-\left[2(t^c_Z\cdot p_{\chi_i})(t^d_Z\cdot p_{\chi_i})+
		{\textstyle \frac{1}{2}}(m_{\chi_i}^2+m_{\chi_n}^2-m_Z^2)
		\delta^{cd}\right]|O^{''L}_{ni}|^2
	\nonumber\\
&&-\delta^{cd}m_{\chi_i}m_{\chi_n}[(Re O^{''L}_{ni})^2 -(Im
		O^{''L}_{ni})^2],\label{cdD1}
\end{eqnarray}
and $^{c}D_1=0$ due to the Majorana character of the neutralinos.
As a consequence of the completeness relation,
Eq.~(\ref{completeness}), the diagonal coefficients are linearly
dependent
\begin{eqnarray}
	^{11}D_1+\,^{22}D_1+\,^{33}D_1&=&-{\textstyle \frac{3}{2}}D_1.
\end{eqnarray}
For large three momentum $p_{\chi_i}$, the $Z$ boson will
mainly be emitted into the forward direction with respect to
$p_{\chi_i}$, i.e. $\hat p_{\chi_i}\approx\hat p_{Z}$, 
with $\hat p=\vec p/|\vec p|$, so that 
$(t^{1,2}_Z\cdot p_{\chi_i}) \approx 0$ in 
Eq.~(\ref{cdD1}). Therefore, for high energies
$^{11}D_1\approx \,^{22}D_1$, and the contributions for the
non-diagonal coefficients $^{cd}D_1 (c \slashed{=}d)$
will be small.

For the second term of $\rho_{D1}$, Eq.~(\ref{rhoD1}), 
which depends on the polarization of the decaying neutralino,
we obtain
\begin{eqnarray}
\Sigma^{a~\mu\nu}_{D1}
\varepsilon_{\mu}^{\lambda_k\ast}\varepsilon_{\nu}^{\lambda'_k}&=&
  \Sigma^{a}_{D1}\delta^{\lambda_k\lambda_k'}
  +\,^c\Sigma^{a}_{D1}(J^c)^{\lambda_k\lambda_k'}
  + \,^{cd}\Sigma^{a}_{D1}(J^{cd})^{\lambda_k\lambda_k'},
\end{eqnarray}
summed over $c$, $d$, with
 \begin{eqnarray}
^c\Sigma^{a}_{D1}&=&\frac{2}{m_Z}\Big\{
	\Big[|O^{''L}_{ni}|^2m_{\chi_i}+[(Re O^{''L}_{ni})^2 -(Im
			O^{''L}_{ni})^2 ]m_{\chi_n}\Big]\nonumber\\ &&\times
	\left[(s^a_{\chi_i}\cdot p_Z)(t^c_Z\cdot p_{\chi_i})-
		(s^a_{\chi_i}\cdot t^c_Z)(p_Z\cdot p_{\chi_i})\right]
	+|O^{''L}_{ni}|^2m_{\chi_i}m_Z^2(s^a_{\chi_i}\cdot t^c_Z)\nonumber\\ &&
	-2(Im O^{''L}_{ni})(Re O^{''L}_{ni})m_{\chi_n}
	\epsilon_{\mu\nu\rho\sigma}s^{a\mu}_{\chi_i}p_{\chi_i}^{\nu}
	p_Z^{\rho}t^{c\sigma}_Z\Big\} \label{csigmaaD1},
\end{eqnarray}
and $\Sigma^{a}_{D1}=\,^{cd}\Sigma^{a}_{D1}=0$ due to the Majorana
character of the neutralinos. A similar expansion for the $Z$ 
decay matrix, Eq.~(\ref{rhoD2}), results in
\begin{eqnarray}
	 \rho_{D2}(Z)_{\lambda'_k\lambda_k}&=&   
D_2 \delta^{\lambda_k'\lambda_k} 
  + \,^cD_2(J^c)^{\lambda_k'\lambda_k}
  + \,^{cd}D_2(J^{cd})^{\lambda_k'\lambda_k},
\end{eqnarray}
where we sum over $c$, $d$, with 
\begin{eqnarray}
D_2&=&{\textstyle\frac{2}{3}}(R_f^2+L_f^2)m_Z^2,\label{D2}\\
^{c}D_2&=&2(R_f^2-L_f^2)m_Z(t^c_Z\cdot p_{\bar f}),\label{cD2}\\
^{cd}D_2&=&(R_f^2+L_f^2)\left[2(t^c_Z\cdot p_{\bar f})(t^d_Z\cdot p_{\bar f})-
	{\textstyle \frac{1}{2}}m_Z^2\delta^{cd}\right].\label{cdD2}
\end{eqnarray}
As a consequence of the completeness relation,
Eq.~(\ref{completeness}), the diagonal coefficients are linearly
dependent
\begin{eqnarray}
	^{11}D_2+\,^{22}D_2+\,^{33}D_2&=&-{\textstyle \frac{3}{2}}D_2.
\end{eqnarray}
For large three-momentum $p_{Z}$, the fermion $\bar f$ will
mainly be emitted into the forward direction with respect to
$p_{Z}$, i.e. $\hat p_{Z}\approx\hat p_{\bar f}$, so that 
$(t^{1,2}_Z\cdot p_{\bar f}) \approx 0$ in 
Eq.~(\ref{cdD2}). Therefore, for high energies
$^{11}D_2\approx \,^{22}D_2$, and the contributions for the
non-diagonal coefficients $^{cd}D_2 (c \slashed{=}d)$ will be small.

Inserting the density matrices (\ref{rhoP}) and (\ref{rhoD1}) 
into Eq.~(\ref{Zdensitymatrix}) leads to:
\begin{eqnarray}\label{Zdensitymatrixunnorm}
\rho_{P}(Z)^{\lambda_k\lambda'_k}&=&
4~|\Delta(\tilde\chi^0_i)|^2~
\left[
	P  D_1 ~\delta^{\lambda_k\lambda_k'}
	+\Sigma_P^a \,^{c}\Sigma^{a}_{D1}~(J^{c})^{\lambda_k\lambda_k'}
	+P  \;^{cd}D_1 ~(J^{cd})^{\lambda_k\lambda_k'}
\right],
\end{eqnarray}
summed over $a,c,d$. Inserting then 
(\ref{Zdensitymatrixunnorm}) and (\ref{rhoD2})
into Eq.~(\ref{amplitude}) leads to:
\begin{equation} \label{amplitude2}
|T|^2 = 4~|\Delta(\tilde{\chi}^{0}_i)|^2~ |\Delta(Z)|^2
		\left[3 P  D_1  D_2 +  
		2\Sigma_P^a \,^c\Sigma_{D1}^a \, ^cD_2 
		+4P( ^{cd} D_1  ^{cd} D_2-
			{\textstyle \frac{1}{3}}\,^{cc} D_1 \, ^{dd} D_2)\right],
\end{equation}
summed over $a,c,d$, which is the decomposition of the amplitude 
squared in its scalar (first term), vector (second term)
and tensor part (third term).

\subsection{$Z$ boson density matrix
     \label{Z boson matrix}}

The polarization of the $Z$ boson, produced in the neutralino decay
(\ref{decay_1}), is given by its $3\times3$ density matrix 
$<\rho(Z)>$ with ${\rm Tr}\{<\rho(Z)>\}=1$.
We obtain $<\rho(Z)>$ in the laboratory system
by integrating Eq.~(\ref{Zdensitymatrixunnorm}) 
over the Lorentz invariant phase space element 
$d{\rm Lips}(s,p_{\chi_j },p_{\chi_n},p_{Z})=
\frac{1}{(2\pi)^2}~d{\rm Lips}(s,p_{\chi_i},p_{\chi_j} )
~d s_{\chi_i} \,\sum_{\pm}
d{\rm Lips}(s_{\chi_i},p_{\chi_n},p_{Z}^{\pm})$,
see Eq.~(\ref{Lips}),
and normalizing by the trace:
\begin{equation}\label{Zdensitymatrixnorm}
<\rho(Z)^{\lambda_k\lambda'_k}>=
\frac{\int \rho_{P}(Z)^{\lambda_k\lambda'_k}~d{\rm Lips}}
		{\int {\rm Tr} \{\rho_{P}(Z)^{\lambda_k\lambda'_k}\}~d{\rm Lips}}
={\textstyle \frac{1}{3}}\delta^{\lambda_k\lambda_k'}
+V_c ~(J^{c})^{\lambda_k\lambda_k'}
+T_{cd}  ~(J^{cd})^{\lambda_k\lambda_k'},\label{defcoef}
\end{equation}
summed over $c$, $d$. The vector and tensor coefficients
$V_c$ and $T_{cd}$ are given by:
\begin{equation}
V_c=\frac{\int |\Delta(\tilde{\chi}^{0}_i)|^2~\Sigma_P^a \,^{c}\Sigma^{a}_{D1} ~d{\rm Lips}}
{3 \int |\Delta(\tilde{\chi}^{0}_i)|^2 ~P  D_1~d{\rm Lips}},\quad
T_{cd}=T_{dc}=
\frac{\int |\Delta(\tilde{\chi}^{0}_i)|^2  ~P  \;^{cd}D_1 ~d{\rm Lips}}
{3 \int |\Delta(\tilde{\chi}^{0}_i)|^2 ~P  D_1~d{\rm Lips}},
\end{equation}
with sum over a. 
The tensor coefficients $T_{12}$ and $T_{23}$ vanish due to
phase-space integration.
The density matrix in the circular basis,
see Eq.~(\ref{circularbasis}), is given  by 
\begin{eqnarray} \label{density1}
	<\rho(Z)^{--}> &= &
	{\textstyle \frac{1}{2}}-V_3+T_{33}, \\
	<\rho(Z)^{00}> &=&-2T_{33},\\
	<\rho(Z)^{-0}> &= &
	{\textstyle \frac{1}{\sqrt{2}}}(V_1+iV_2)-\sqrt{2}\,T_{13},\\
		<\rho(Z)^{-+}> &= & T_{11},\\
	<\rho(Z)^{0+}> &= & {\textstyle \frac{1}{\sqrt{2}}}(V_1+iV_2)
	+\sqrt{2}\,T_{13},\label{density5}
\end{eqnarray}
where we have used $T_{11}+T_{22}+T_{33}=-\frac{1}{2}$
and $T_{12}=T_{23}=0$.

\section{T odd asymmetry
	\label{T odd asymmetry}}

From Eq.~(\ref{amplitude2}) 
one obtains for the asymmetry, Eq.~(\ref{AT1}):
\begin{eqnarray} \label{asym}
	 {\mathcal A}_{f} 
	 = \frac{\int {\rm Sign}[{\mathcal T}_{f}]
		 |T|^2 d{\rm Lips}}
           {\int |T|^2 d{\rm Lips}}
= \frac{\int |\Delta(\tilde{\chi}^{0}_i)|^2 |\Delta(Z)|^2~
	{\rm Sign}[{\mathcal T}_{f}]
		2\Sigma_P^a \, ^c\Sigma_{D1}^a \,^cD_2 d{\rm Lips}}
	{\int |\Delta(\tilde{\chi}^{0}_i)|^2 |\Delta(Z)|^2~
		3 P D_1D_2 d{\rm Lips}},
\end{eqnarray}
summed over $a$, $c$. In the numerator only the vector part of $|T|^2$ 
remains because only the vector
part contains the triple product\footnote
{
Note that if one would replace the triple product 
${\mathcal T}_{f}$ by 
${\mathcal T}_{f} =\vec p_{e^-}\cdot(\vec p_{\chi_i} \times \vec p_{Z})$,
and would calculate the corresponding asymmetry,
where the $Z$ boson polarization is summed, all spin correlations and 
thus this asymmetry would vanish identically because 
of the Majorana properties of the neutralinos.
}
${\mathcal T}_{f} = 
 \vec p_{e^-}\cdot(\vec p_{f} \times \vec p_{\bar f})$.
In the denominator the vector and tensor parts of $|T|^2$ vanish,
because for complete phase space integrations the spin correlations 
are eliminated. Due to the correlations between
the $\tilde\chi^0_i$ and the $Z$ boson polarization, 
$\Sigma_P^a \, ^c\Sigma_{D1}^a$, there are 
CP odd contributions to the asymmetry ${\mathcal A}_{f}$ which stem
from the neutralino production process, see Eq.~(\ref{production}),
and/or from the neutralino decay process, see Eq.~(\ref{decay_1}).
The contribution from the production is given by the term with $a=2$ 
in Eq.~(\ref{asym}) and it is proportional to $\Sigma^{2}_P$, 
Eq.~(\ref{rhoP}), which is the transverse polarization of the 
neutralino perpendicular to the production plane. 
For $e^+e^- \to\tilde\chi^0_i \tilde\chi^0_i$ we have $\Sigma^{2}_P=0$.
The contributions  from the decay, which are the terms with $a=1,3$
in Eq.~(\ref{asym}), are proportional to  
\begin{eqnarray}\label{adependence}
	^c\Sigma_{D1}^a \,^cD_2&\supset&
	-8m_{\chi_n}
	(Im O^{''L}_{ni})(Re O^{''L}_{ni})(R_f^2-L_f^2)
	(t^c_Z\cdot p_{\bar f})
	\epsilon_{\mu\nu\rho\sigma}s^{a\mu}_{\chi_i}p_{\chi_i}^{\nu}
	p_Z^{\rho}t^{c\sigma}_Z,
\end{eqnarray}
see last term of Eq.~(\ref{csigmaaD1}), 
which contains the $\epsilon$-tensor.
Thus ${\mathcal A}_{f}$ can be enhanced (reduced) if the contributions 
from production and decay  have the same (opposite) sign.
Note that the contributions from the decay would vanish 
for a two-body decay of the neutralino into a scalar particle.
In this case the remaining contributions from the
production are multiplied by a decay factor 
$\propto (|R|^2-|L|^2)$ \cite{olaf1}, and thus  
${\mathcal A}_{f}\propto (|R|^2-|L|^2)/(|R|^2+|L|^2)$,
where $R$ and $L$ are the right and left couplings
of the scalar particle to the neutralino.

For the measurement of ${\mathcal A}_{f}$ 
the charges and the flavors of $f$ and $\bar f$
have to be  distinguished. For $f=e,\mu$ this will be 
possible on an event by event basis. 
For $f=\tau$ it will be possible after taking into account
corrections due to the reconstruction of the $\tau$ 
momentum. For $f=q$ the distinction of the quark 
flavors should be possible by flavor tagging in the 
case $q=b,c$ \cite{Damerell:1996sv}. 
However, in this case the quark charges will  be distinguished
statistically for a given event sample only  \cite{Aubert:2002rg}.
Note that ${\mathcal A}_{q}$ is always larger than
${\mathcal A}_{\ell}$, due to the dependence of ${\mathcal A}_{f}$ 
on the $Z$-$\bar f$-$f$ couplings \cite{oshimo,staudecay}:
\begin{eqnarray} \label{propto}
{\mathcal A}_{f}\propto \frac{R_f^2-L_f^2}{R_f^2+L_f^2}
\quad\Rightarrow
{\mathcal A}_{b(c)}=
\frac{R_{\ell}^2+L_{\ell}^2}
     {R_{\ell}^2-L_{\ell}^2}
\frac{R_{b(c)}^2-L_{b(c)}^2}
     {R_{b(c)}^2+L_{b(c)}^2}~
	  {\mathcal A}_{\ell}\simeq 
	6.3~(4.5)\times{\mathcal A}_{\ell},
\end{eqnarray}
which follows from Eqs.~(\ref{D2}), (\ref{cD2}) and (\ref{asym}).

The relative statistical error of
${\mathcal A}_{f}$ is given by $\delta {\mathcal A}_{f} = 
\Delta {\mathcal A}_{f}/|{\mathcal A}_{f}| = 
S_f/(|{\mathcal A}_{f}| \sqrt{N})$ \cite{olaf1},
with $S_f$ standard deviations
and $N={\mathcal L} \cdot\sigma_t$ the number of events with 
${\mathcal L}$ the integrated luminosity and 
the cross section 
$\sigma_t=\sigma(e^+e^-\to\tilde\chi^0_i\tilde\chi^0_j) 
\times{\rm BR}(\tilde{\chi}^0_i \to Z\tilde{\chi}_n^0)\times
{\rm BR}(Z\to f\bar f)$. Taking $\delta {\mathcal A}_{f}
=1$ it follows 
$S_f = |{\mathcal A}_{f}| \sqrt{N}$.
Note that $S_f$ is larger  for $f=b,c$ than for $f=\ell=e,\mu,\tau$ with
$S_{b} \simeq 7.7\times S_{\ell}$ and $S_{c} \simeq 4.9\times S_{\ell}$,
which follows from Eq.~(\ref{propto}) and from
${\rm BR}(Z\to b\bar b) \simeq1.5\times{\rm BR}(Z\to\ell\bar\ell)$,
${\rm BR}(Z\to c\bar c) \simeq1.2\times{\rm BR}(Z\to\ell\bar\ell)$.

\section{Numerical results
	\label{Numerical results}}

We present numerical results for the  
$Z$ density matrix $<\rho(Z)>$, Eq.~(\ref{Zdensitymatrixnorm}),
the asymmetry  ${\mathcal A}_{\ell} (\ell=e,\mu,\tau)$, 
Eq.~(\ref{AT1}), and the cross section 
$\sigma_{t}=\sigma(e^+e^-\to\tilde\chi^0_i\tilde\chi^0_j ) \times
{\rm BR}( \tilde\chi^0_i \to \tilde\chi^0_1 Z)\times
{\rm BR}(Z \to \ell \bar \ell)$.
For the branching ratio $Z\to\ell\bar\ell$, summed over
$\ell=e,\mu,\tau$, we take the experimental  value 
${\rm BR}(Z\to\ell\bar\ell)=0.1$ \cite{PDG}.
The values for ${\mathcal A}_{b,c}$ may be obtained
from  Eq.~(\ref{propto}).
We choose  a center  of mass energy of   $\sqrt{s} = 800$ GeV
and longitudinally polarized beams with
beam polarizations $(P_{e^-},P_{e^+})=(\pm0.8,\mp0.6)$.
We study the dependence of $<\rho(Z)>$, ${\mathcal A}_{\ell}$ and
$\sigma_{t}$ on the MSSM parameters 
$\mu = |\mu| \, e^{ i\,\varphi_{\mu}}$ and 
$M_1 = |M_1| \, e^{ i\,\varphi_{M_1}}$. 
For all scenarios we keep $\tan \beta=10$. 
In order to reduce the number of parameters, we assume the 
relation $|M_1|=5/3  M_2\tan^2\theta_W $ and  
use the renormalization group equations \cite{hall} for the 
selectron and smuon masses,
$m_{\tilde\ell_R}^2 = m_0^2 +0.23 M_2^2
-m_Z^2\cos 2 \beta \sin^2 \theta_W$, 
$m_{\tilde\ell_L  }^2 = m_0^2 +0.79 M_2^2
+m_Z^2\cos 2 \beta(-1/2+ \sin^2 \theta_W)$,
taking $m_0=300$ GeV.

Our numerical results presented below are obtained at tree level.
One-loop corrections to $e^+e^-\to\tilde\chi^0_i \tilde\chi^0_j$
have been given in \cite{HEPHY} for real MSSM parameters. They are
of the order of a few percent and may reach values up to 10\%. As
the bulk of the one-loop corrections are presumably CP-even, we
expect that they will not significantly change our tree-level result
for ${\mathcal A}_{f}$. For an appropriate analysis of the one-loop
corrections to ${\mathcal A}_{f}$ it would be necessary to adopt
the formulae of  \cite{HEPHY} to the case of complex MSSM
parameters, which is beyond the scope of the present paper.

The experimental upper limits on the electric dipole moments (EDMs)
of electron and neutron may restrict the phases 
$\varphi_{\mu}$ and $\varphi_{M_1}$. These restrictions are very
model dependent. They are less severe when cancellations between
the contributions of different SUSY phases occur. For example, in
the constrained MSSM  the phase $\varphi_{\mu}$ is restricted to 
$|\varphi_{\mu}|\lsim0.1\pi$, whereas the phase $\varphi_{M_1}$ is
not restricted, but correlated with $\varphi_{\mu}$ \cite{Ibrahim}.
In most of our numerical examples below we have chosen 
$\varphi_{M_1}=\pm \pi/2$, $\varphi_{\mu}=0$, which agrees with the
constraints from the electron and neutron EDMs. In order to show
the full phase dependences of the CP asymmetry ${\mathcal A}_{f}$,
in one example we study its $\varphi_{\mu}$ behavior in the whole 
$\varphi_{\mu}$ range, relaxing in this case the restrictions from
the EDMs. However, as shown in \cite{BMPW}, if also lepton flavor
violating terms are included, the EDM constraints on $\varphi_{\mu}$
disappear.

For the calculation of the neutralino widths $\Gamma_{\chi_i}$ and 
the branching ratios ${\rm BR}( \tilde\chi^0_i \to \tilde\chi^0_1 Z)$ 
we neglect three-body decays and 
include the following two-body decays, if kinematically allowed,
\begin{eqnarray}
	\tilde\chi^0_i &\to& \tilde e_{R,L} e,~ 
	\tilde \mu_{R,L}\mu,~
	\tilde\tau_{m}\tau,~
	\tilde\nu_{\ell} \bar\nu_{\ell},~
	\tilde\chi^0_n Z,~
	\tilde\chi^{\mp}_m W^{\pm},~
	\tilde\chi^0_n H_1^0,~
	\ell=e,\mu,\tau, ~ m=1,2,~ n<i
\end{eqnarray}
with $H_1^0$ being the lightest neutral Higgs boson.
The Higgs parameter is chosen $m_{A}=1000$~GeV and thus 
the decays  $\tilde\chi^0_i \to \tilde\chi^{\pm}_n H^{\mp}$
into the the charged Higgs bosons,
and the decays $\tilde\chi^0_i \to \tilde\chi^0_n~H_{2,3}^0$
into  the heavy neutral Higgs bosons are forbidden in our scenarios.
In the stau sector, we fix the trilinear scalar coupling
parameter $A_{\tau}=250$ GeV.

\subsection{Production of $\tilde\chi^0_1 \, \tilde\chi^0_2$ }

\begin{figure}[h]
 \begin{picture}(20,20)(0,-2)
	\put(2.2,16.5){\fbox{$\sigma(e^+e^- \to\tilde{\chi}^0_1 
			\tilde{\chi}^0_2)$ in fb}}
	\put(0,9){\includegraphics{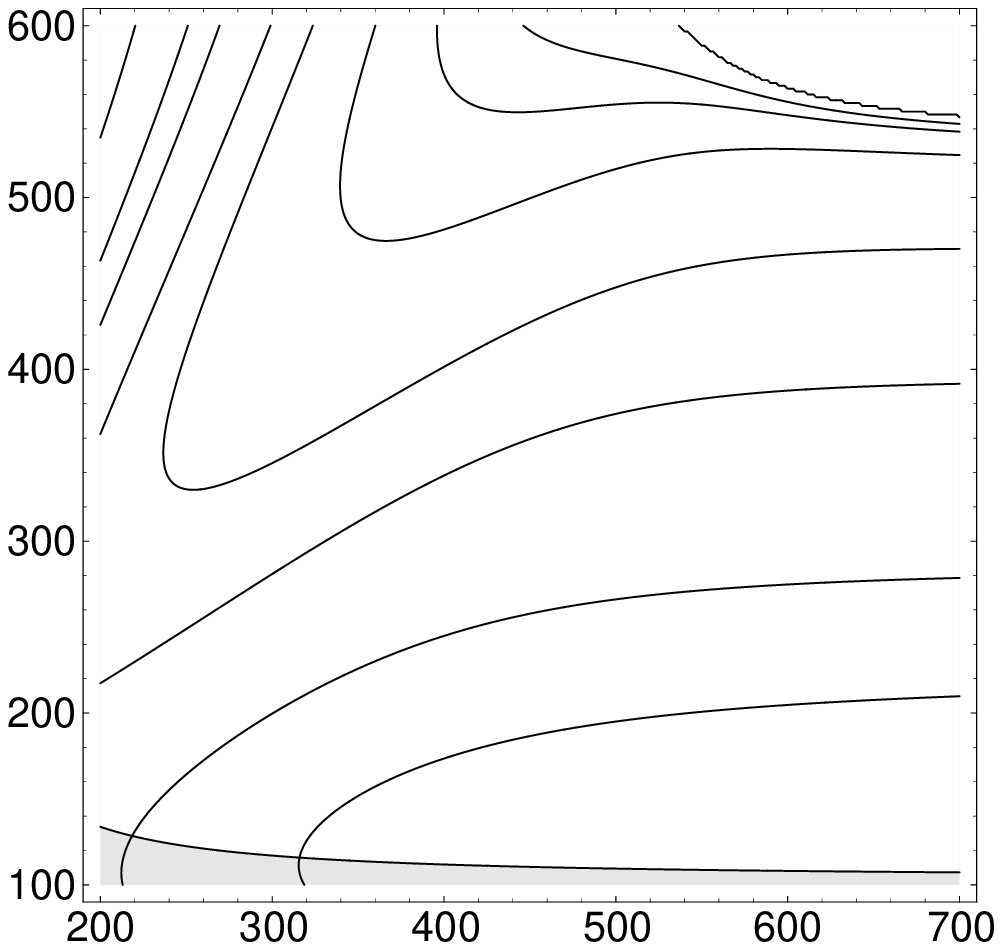}}
   \put(5.5,8.8){$|\mu|$~/GeV}
   \put(0,16.3){$M_2$~/GeV }
	\put(6.5,15.6){\begin{picture}(1,1)(0,0)
			\CArc(0,0)(6,0,380)
			\Text(0,0)[c]{{\scriptsize A}}
		\end{picture}}
	\put(4.3,15.6){ \footnotesize$10$}
	\put(3.3,15.3){ \footnotesize$15$}
	\put(2.6,14.5){ \footnotesize$25$}
	\put(2.8,13.5){ \footnotesize$50$}
	\put(3.5,13.){ \footnotesize$75$}
	\put(4.,11.9){ \footnotesize$100$}
	\put(5.3,11.1){ \footnotesize$110$}
\put(0.5,8.8){Fig.~\ref{plot_12}a}
   \put(8,9){\includegraphics{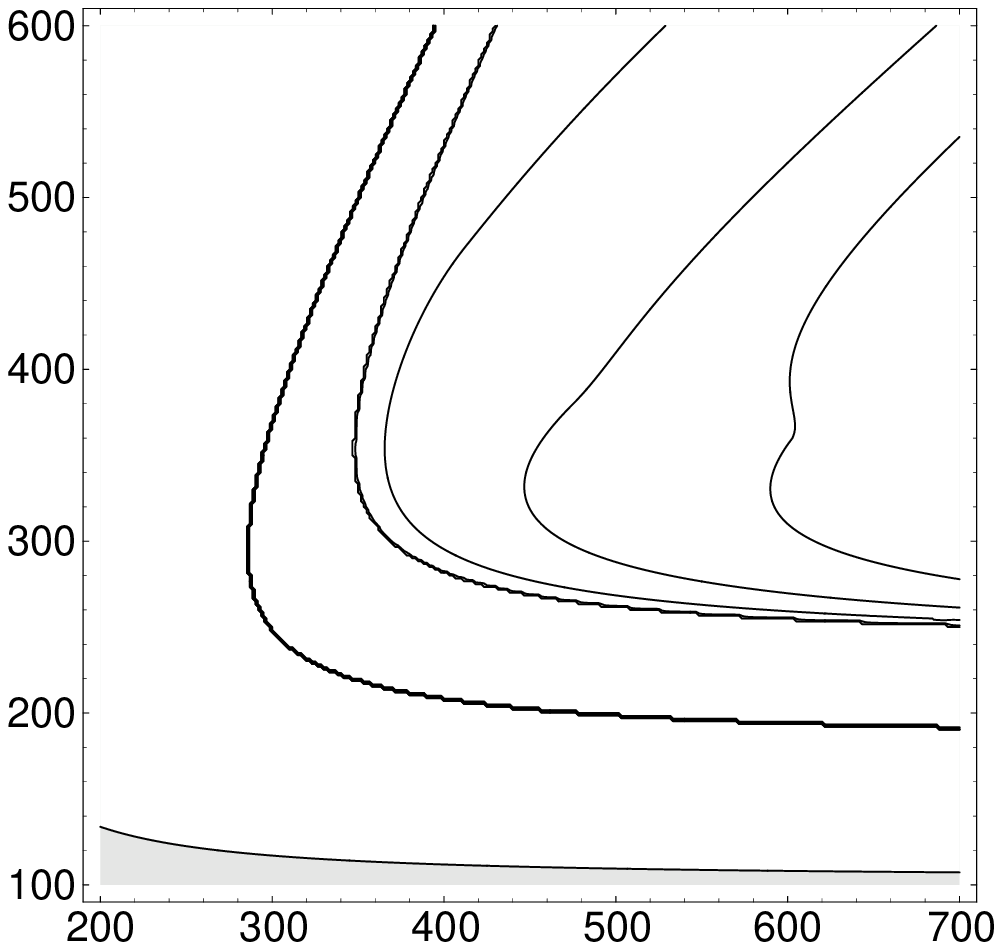}}
	\put(10.2,16.5){\fbox{BR$(\tilde{\chi}^0_2 \to Z\tilde{\chi}_1^0)$ in \%}}
   \put(13.5,8.8){$|\mu|$~/GeV}
	\put(8,16.3){$M_2$~/GeV }
	\put(9.3,15){\begin{picture}(1,1)(0,0)
			\CArc(0,0)(6,0,380)
			\Text(0,0)[c]{{\scriptsize B}}
	\end{picture}}
	\put(10.4,11.8){\begin{picture}(1,1)(0,0)
			\CArc(0,0)(6,0,380)
			\Text(0,0)[c]{{\scriptsize C}}
	\end{picture}}
	\put(13.6,12.7){ \footnotesize 15}
	\put(12.2,13.1){ \footnotesize 20}
	\put(11.,13.5){ \footnotesize 30}
	\put(11.,11.5){ \footnotesize 100}
\put(8.5,8.8){Fig.~\ref{plot_12}b}
	\put(0,0){\includegraphics{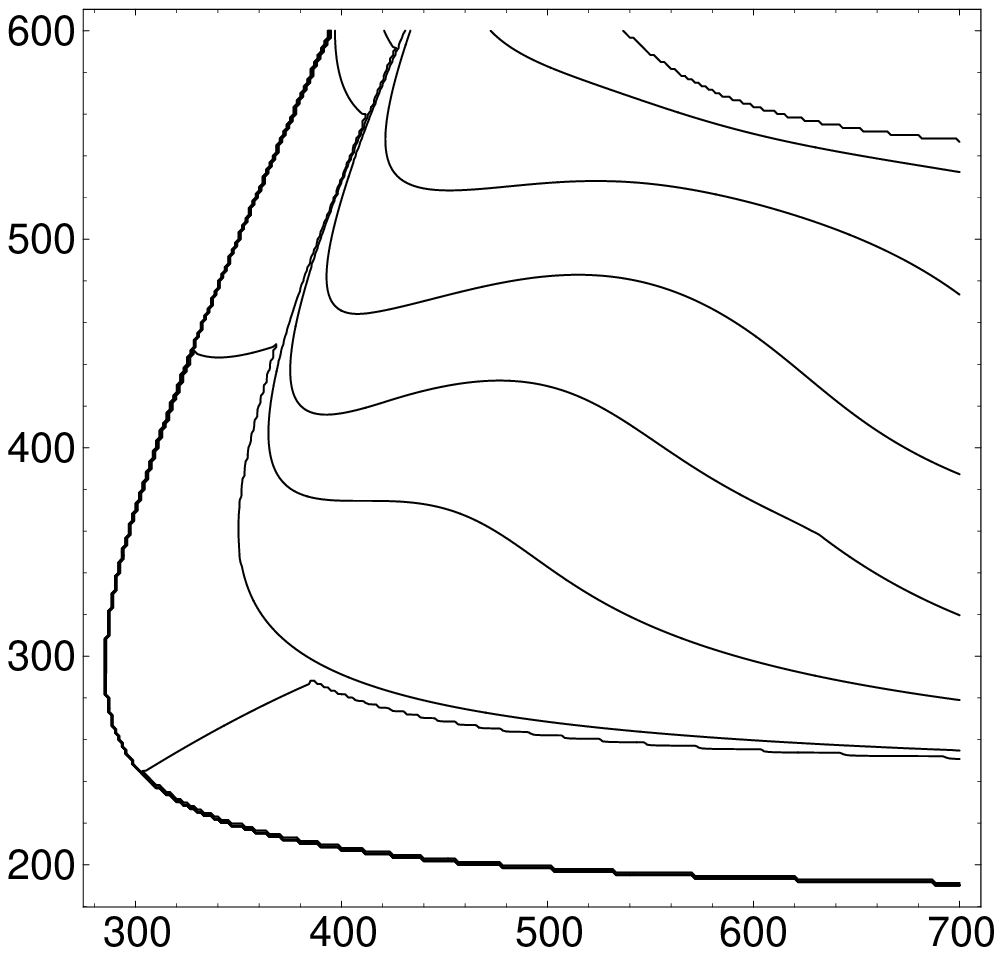}}
	\put(3.5,7.5){\fbox{$\sigma_{t}$ in fb}}
	\put(5.5,-0.3){$|\mu|$~/GeV}
	\put(0,7.3){$M_2$~/GeV }
	\put(6.5,6.6){\begin{picture}(1,1)(0,0)
			\CArc(0,0)(6,0,380)
			\Text(0,0)[c]{{\scriptsize A}}
	\end{picture}}
		\put(1.3,6){\begin{picture}(1,1)(0,0)
			\CArc(0,0)(6,0,380)
			\Text(0,0)[c]{{\scriptsize B}}
		\end{picture}}
	\put(4.5,5.9){\footnotesize 0.3}
	\put(4.3,5.4){\footnotesize 0.6}
	\put(3.8,4.7){\footnotesize 0.9}
	\put(3.3,4.){\footnotesize 1.2}
	\put(2.9,3.){\footnotesize 1.5}
	\put(2.1,2.6){\footnotesize 3}
	\put(1.8,1.65){\footnotesize 6}
	\put(2.15,6.1){\scriptsize 1.5}
	\put(1.5,4.15){\footnotesize 3}
	\put(0.5,-0.3){Fig.~\ref{plot_12}c}
   \put(8,0){\includegraphics{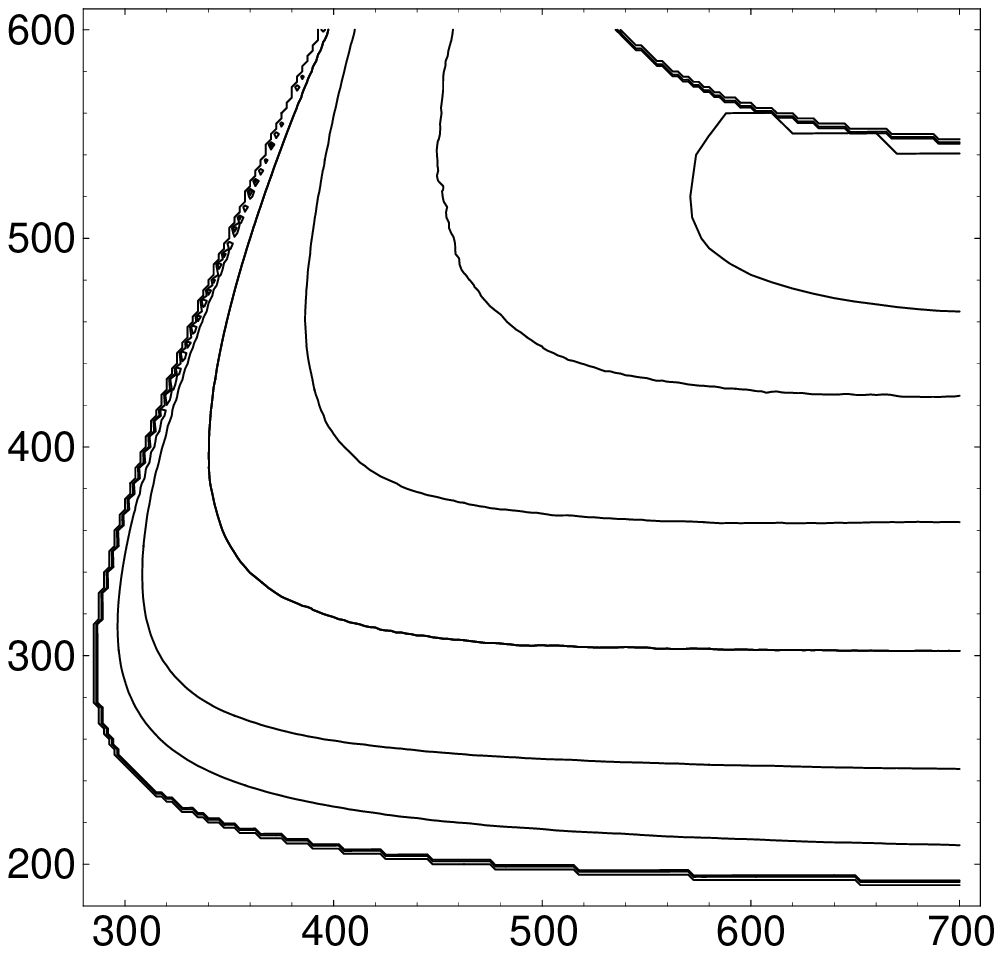}}
	\put(11.5,7.5){\fbox{${\mathcal A}_{\ell}$ in \% }}
	\put(8,7.3){$M_2$~/GeV }
	\put(13.5,-0.3){$|\mu|$~/GeV}
	\put(14.5,6.6){\begin{picture}(1,1)(0,0)
			\CArc(0,0)(6,0,380)
			\Text(0,0)[c]{{\scriptsize A}}
	\end{picture}}
		\put(9.3,6){\begin{picture}(1,1)(0,0)
			\CArc(0,0)(6,0,380)
			\Text(0,0)[c]{{\scriptsize B}}
	\end{picture}}
	\put(13.1,5.5){\footnotesize -1.6}
	\put(12.1,4.8){\footnotesize -1.5}
	\put(11.3,3.7){\footnotesize -1}
	\put(10.6,2.7){\footnotesize 0}
	\put(10.1,1.9){\footnotesize 1}
	\put(10.,1.35){\footnotesize 1.5}
   \put(8.5,-0.3){Fig.~\ref{plot_12}d}
\end{picture}
\vspace*{-1.5cm}
\caption{
	Contour plots for  
	\ref{plot_12}a: $\sigma(e^+e^- \to\tilde{\chi}^0_1\tilde{\chi}^0_2)$, 
	\ref{plot_12}b: BR$(\tilde{\chi}^0_2 \to Z\tilde{\chi}_1^0)$,
	\ref{plot_12}c: $\sigma_t=\sigma(e^+e^-\to\tilde\chi^0_1\tilde\chi^0_2 ) 
	\times{\rm BR}(\tilde{\chi}^0_2 \to Z\tilde{\chi}_1^0)\times
	{\rm BR}(Z\to\ell\bar\ell)$ with BR($Z\to\ell\bar\ell$) = 0.1,
	\ref{plot_12}d: the asymmetry ${\mathcal A}_{\ell}$,
	in the $|\mu|$--$M_2$ plane for $\varphi_{M_1}=0.5\pi $, 
	$\varphi_{\mu}=0$, taking  $\tan \beta=10$, $m_0=300$ GeV,
	$\sqrt{s}=800$ GeV and $(P_{e^-},P_{e^+})=(-0.8,0.6)$.
	The area A (B) is kinematically forbidden by
	$m_{\tilde\chi^0_1}+m_{\tilde\chi^0_2}>\sqrt{s}$
	$(m_{Z}+m_{\tilde\chi^0_1}> m_{\tilde\chi^0_2})$.
	In area C of plot \ref{plot_12}b:  
	BR$(\tilde{\chi}^0_2 \to Z\tilde{\chi}_1^0)=100 \%$.
	The gray  area is excluded by $m_{\tilde\chi_1^{\pm}}<104 $ GeV. 
	\label{plot_12}}
\end{figure}

In Fig.~\ref{plot_12}a we show the cross section for
$\tilde\chi^0_1 \tilde\chi^0_2$ production in the $|\mu|$--$M_2$
plane for $\varphi_{\mu}=0$ and $\varphi_{M_1}=0.5\pi$. 
For $|\mu|  \gsim 250 $ GeV the left selectron exchange dominates
due to the larger $\tilde\chi^0_2-\tilde e_L$ coupling,
so that the choice of polarization $(P_{e^-},P_{e^+})=(-0.8,0.6)$
enhances the cross section, which reaches values of more than 110 fb. 
The branching ratio  
${\rm BR} (\tilde\chi^0_2 \to Z \tilde\chi^0_1)$
is shown in Fig.~\ref{plot_12}b. 
The branching ratio can even be 100\% and decreases
with increasing $|\mu|$ and $M_2$, when the two-body decays
into sleptons and/or into the lightest neutral Higgs boson   
are kinematically allowed. 
The cross section 
$\sigma_t=\sigma(e^+e^-\to\tilde\chi^0_1\tilde\chi^0_2 ) 
\times{\rm BR}(\tilde{\chi}^0_2 \to Z\tilde{\chi}_1^0)\times
{\rm BR}(Z\to\ell\bar\ell)$ is shown in Fig.~\ref{plot_12}c.
Due to the small branching ratio  BR($Z\to\ell\bar\ell$) = 0.1,
$\sigma_t$ does not exceed 7 fb.
Fig.~\ref{plot_12}d shows the $|\mu|$--$M_2$ dependence 
of the asymmetry ${\mathcal A}_{\ell}$ for 
$\varphi_{M_1}=0.5\pi $ and $\varphi_{\mu}=0$.
The asymmetry $|{\mathcal A}_{\ell}|$ can reach a value of $1.6 \% $.
On the contour 0 in Fig.~\ref{plot_12}d, the (positive) contributions
from the production cancel the (negative) contributions from the
decay.
We also studied the $\varphi_{\mu}$ dependence of 
${\mathcal A}_{\ell}$. In the $|\mu|$--$M_2$ plane for $\varphi_{M_1}=0$ 
and $\varphi_{\mu}=0.5\pi$ we found $|{\mathcal A}_{\ell}|<0.5\%$. 

\begin{figure}[h]
\setlength{\unitlength}{1cm}
			\begin{minipage}{0.45\textwidth}
 \begin{picture}(10,6.5)(.5,0)
	\put(0,0){\includegraphics{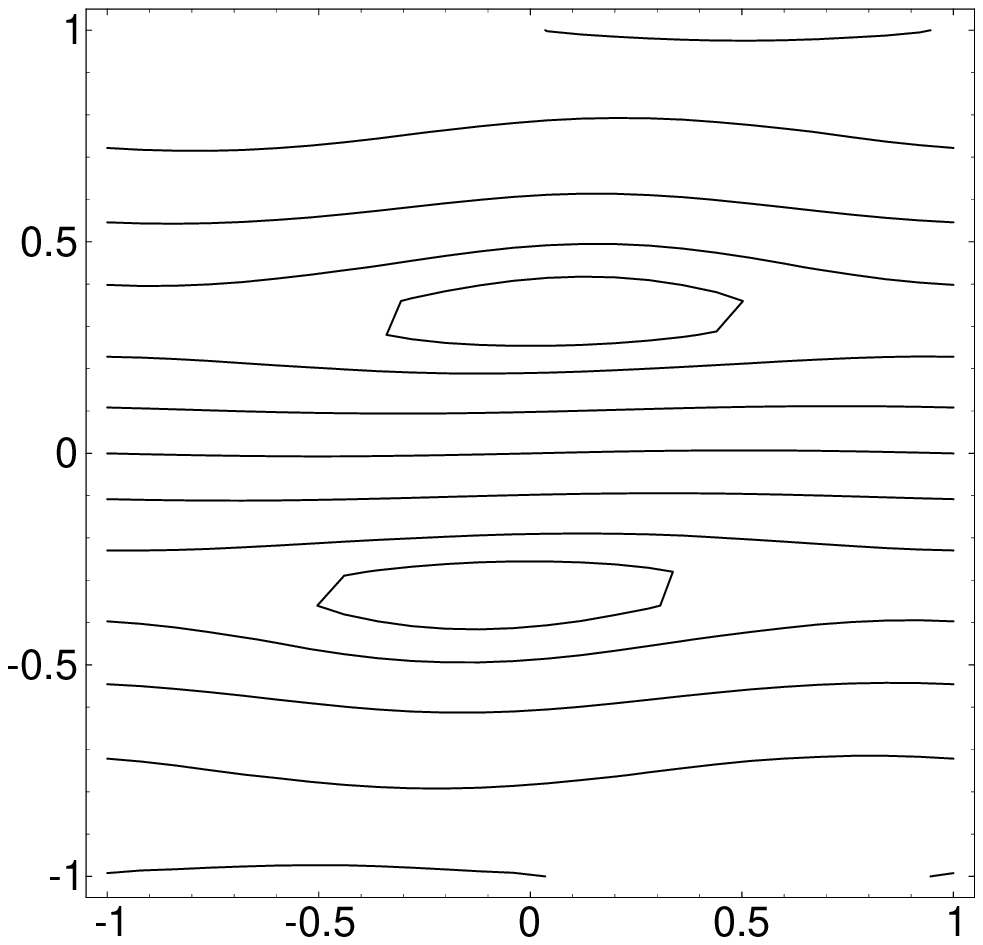}}
	\put(3.3,7.5){\fbox{${\mathcal A}_{\ell}$ in \% }}
	\put(6.5,-.3){$\varphi_{\mu}/\pi$}
	\put(0.3,7.3){$ \varphi_{M_1}/\pi$ }
		\put(6.2,6.45){\scriptsize 0}
		\put(5.8,6.15){\scriptsize 0}
		\put(5.9,5.6){\scriptsize 0.7}
		\put(5.6,5.2){\scriptsize 1.2}
		\put(4.8,4.7){\scriptsize 1.4}
		\put(5.8,4.5){\scriptsize 1.2}
		\put(6.1,4.15){\scriptsize 0.7}
		\put(6.4,3.8){\scriptsize 0}
		\put(5.7,3.5){\scriptsize -0.7}
		\put(5.4,3.2){\scriptsize -1.2}
		\put(4.3,2.7){\scriptsize -1.4}
		\put(5.3,2.6){\scriptsize -1.2}
		\put(5.6,2.15){\scriptsize -0.7}
		\put(6.0,1.65){\scriptsize 0}
		\put(2.5,0.85){\scriptsize 0}
	\end{picture}
\caption{Contour lines of the asymmetry ${\mathcal A}_{\ell}$
for $e^+e^-\to\tilde\chi^0_1\tilde\chi^0_2; 
	  \tilde{\chi}^0_2 \to Z\tilde{\chi}_1^0;
	Z\to\ell\bar\ell (\ell=e,\mu,\tau)$,	
	in the $\varphi_{\mu}$--$\varphi_{M_1}$ plane
for $M_2=250$ GeV and $|\mu|=400$ GeV, taking
$\tan \beta=10$, $m_0=300$ GeV,
$\sqrt{s}=800$ GeV and $(P_{e^-},P_{e^+})=(-0.8,0.6)$.
\label{plot_3}}
\end{minipage}
\hspace*{0.5cm}
%
	\begin{minipage}{0.45\textwidth}
 \begin{picture}(10,6.5)(.3,0)
   \put(-1,8){\includegraphics{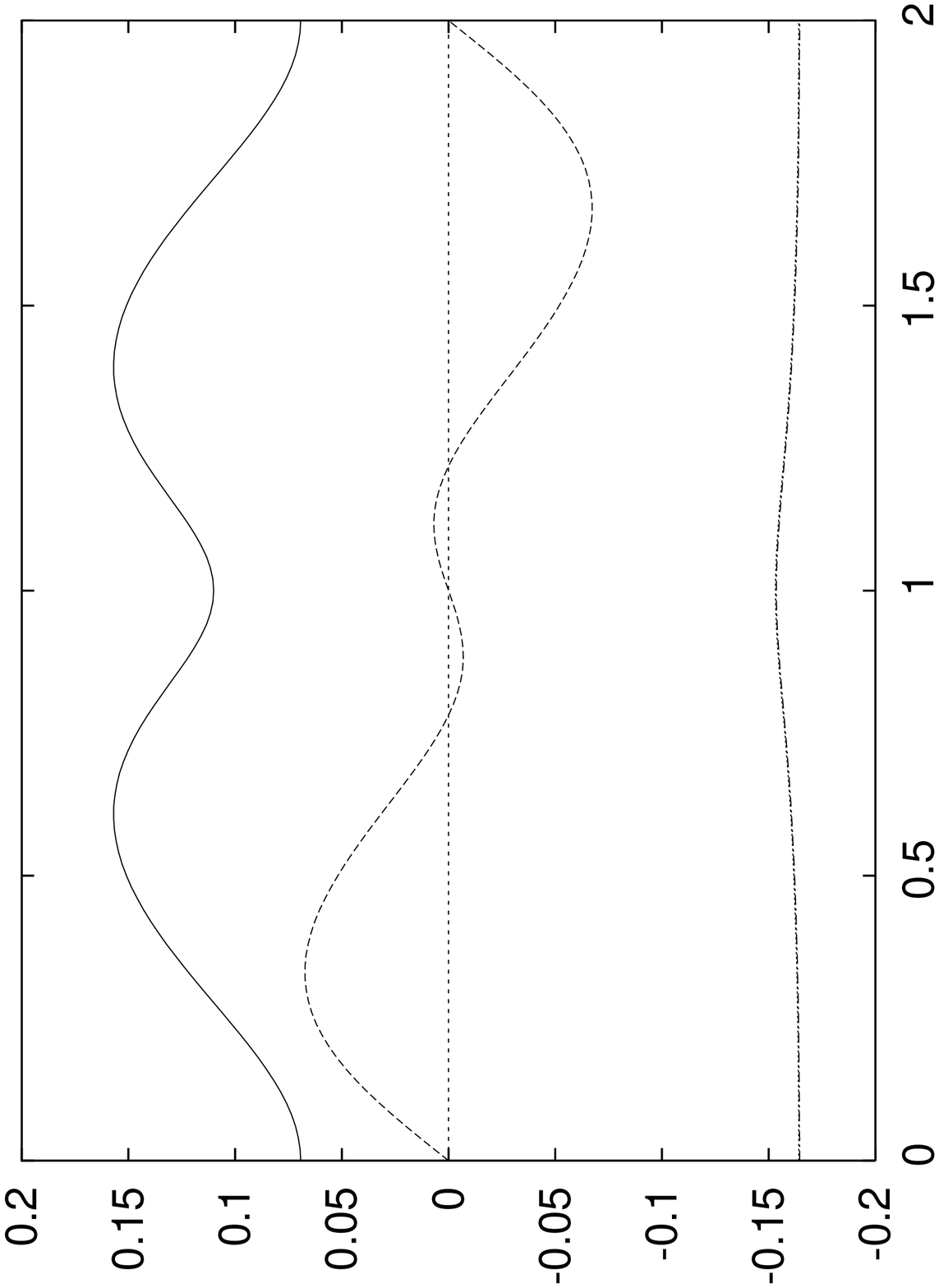}}
	\put(2.5,7.5){\fbox{ matrix elements}}
	\put(6.3,-0.3){$\varphi_{M_1}/\pi $}
	\put(3.5,5.85){\footnotesize $V_1 $}
	\put(2.6,4.5){\footnotesize $ V_2$}
	\put(1.5,3.35){\footnotesize $ V_3\approx 0$}
	\put(3.3,1.65){\footnotesize $ T_{11}\approx T_{22}$}
\end{picture}
\caption{
	Dependence on $\varphi_{M_1}$ of the 
	vector $(V_i)$ and tensor $(T_{ii})$ 
	elements of the $Z$ density matrix  $<\rho(Z)>$,
	for $e^+e^-\to\tilde\chi^0_1\tilde\chi^0_2;
	\tilde{\chi}^0_2 \to Z\tilde{\chi}_1^0$, 
	for $M_2=250$ GeV and $|\mu|=400$ GeV, 
	taking $\varphi_{\mu}=0$, $\tan \beta=10$, $m_0=300$ GeV,
	$\sqrt{s}=800$ GeV 
	and $(P_{e^-},P_{e^+})=(-0.8,0.6)$.
\label{plot_4}}
\end{minipage}
\end{figure}

In Fig.~\ref{plot_3} we show the 
$\varphi_{\mu}$--$\varphi_{M_1}$ dependence of 
${\mathcal A}_{\ell}$ for 
$|\mu|=400$ GeV and $M_2=250$ GeV. The value
of ${\mathcal A}_{\ell}$ depends stronger on $\varphi_{M_1}$
than on $\varphi_{\mu}$. 
It is remarkable
that the maximal phases of  $\varphi_{M_1},\varphi_{\mu}=\pm \pi/2$ do
not  lead to the highest values of 
${\mathcal A}_{\ell} \approx \pm 1.4\%$, which are reached for
$(\varphi_{M_1},\varphi_{\mu}) \approx (\pm0.3 \pi,0)$.
The reason for this is that the spin-correlation terms 
$\Sigma_P^a \,^c\Sigma_{D1}^a \,^cD_2$ in the numerator of 
${\mathcal A}_{f}$, Eq.~(\ref{asym}), are products of CP odd and
CP even factors. The CP odd (CP even) factors have a sine-like
(cosine-like) phase dependence. Therefore, the maximum of the CP
asymmetry ${\mathcal A}_{f}$ is shifted from 
$\varphi_{M_1},\varphi_{\mu}= \pm \pi/2$ to a smaller or larger
value.

In the $\varphi_{\mu}$--$\varphi_{M_1}$ region shown 
in Fig.~\ref{plot_3} also the cross section 
$\sigma_t=\sigma(e^+e^-\to\tilde\chi^0_1\tilde\chi^0_2 ) 
\times{\rm BR}(\tilde{\chi}^0_2 \to Z\tilde{\chi}_1^0)\times
{\rm BR}(Z\to\ell\bar\ell)$ with 
${\rm BR}(\tilde{\chi}^0_2 \to Z\tilde{\chi}_1^0)=1$
and BR($Z\to\ell\bar\ell$) = 0.1,
is rather insensitive to $\varphi_{\mu}$ and 
ranges between 7 fb ($\varphi_{M_1}=0$) and 14 fb ($\varphi_{M_1}=\pm\pi$).
For the leptonic decay of the $Z$, the standard   deviations
are given by $S_{\ell} =|{\mathcal A}_{\ell}| 
\sqrt{{\mathcal L}\cdot\sigma_t}$,
and for the hadronic decays by 
$S_{b(c)}=7.7(4.9)S_{\ell}$, see Section~\ref{T odd asymmetry}.
For ${\mathcal L}= 500$ fb$^{-1}$ and 
$(\varphi_{M_1},\varphi_{\mu})=(\pm0.3 \pi,0)$ in 
Fig.~\ref{plot_3} we find 
$S_{b(c)}=8(5)$ and thus ${\mathcal A}_{b(c)}$ could be measured.
However note that we have $S_{\ell}<1$ in this scenario and thus
${\mathcal A}_{\ell}$ cannot be measured at the 
68\% confidence level $(S_{\ell}=1)$.
In Fig.~\ref{plot_4} we show the 
$\varphi_{M_1}$ dependence of the vector $(V_i)$ and tensor $(T_{ii})$ 
elements of the $Z$ density matrix  $<\rho(Z)>$.
The elements $T_{11}$, $T_{22}$ and $V_1$ have a CP even behavior.
The element $V_2$ is CP odd and is not only zero at 
$\varphi_{M_1}=0$ and $\varphi_{M_1}=\pi$, 
but also at $\varphi_{M_1}\approx(1\pm0.2)\pi$, 
which is due to the destructive interference of the
contributions from CP violation in production and decay.
The interference of the contributions from the
CP even effects in production and decay cause the two
maxima of $V_1$. As discussed in Section~\ref{Amplitude squared},
the tensor elements $T_{11}$ and $T_{22}$ are almost 
equal. Compared to  $V_1$ and $V_2$, they have the same order of 
magnitude but their dependence on $\varphi_{M_1}$ is rather weak.
Furthermore, the other elements are small, i.e. $T_{13}, V_3<10^{-6}$
and thus the density matrix  $<\rho(Z)>$ 
assumes a symmetric shape. In the CP conserving case, e.g. for
$\varphi_{M_1}=\varphi_{\mu}=0$,
$M_2=250$ GeV,  $|\mu|=400$ GeV, 
$\tan \beta=10$, $m_0=300$ GeV, 
$\sqrt{s}=800$ GeV and  $(P_{e^-},P_{e^+})=(-0.8,0.6)$
it reads:
\begin{eqnarray}
<\rho(Z)>  =\left(
        \begin{array}{ccc}
			  0.329  & 0.049 & 0.0003\\
			  0.049  & 0.343 & 0.049 \\
			  0.0003 & 0.049 & 0.329
        \end{array}
	\right).
\end{eqnarray}
In the CP violating case, e.g. for $\varphi_{M_1}=0.5\pi $ 
and the other parameters as above, $<\rho(Z)>$ has 
imaginary parts due to a non-vanishing $V_2$:
\begin{eqnarray}
<\rho(Z)>  =\left(
        \begin{array}{ccc}
			  0.324        & 0.107+  0.037i&0.0003\\
			  0.107 -0.037i& 0.352&0.107 + 0.037i \\
			  0.0003       & 0.107 -0.037i&0.324
        \end{array}
	\right).
\end{eqnarray}
Imaginary parts of $<\rho(Z)>$
are thus an indication of CP violation. 
Note that also the diagonal elements, being CP
even quantities, are changed for
$ \varphi_{M_1}\slashed =0 $ and 
$ \varphi_{\mu}\slashed =0 $. This fact has been 
exploited in \cite{choi2} as a possibility to determine
the CP violating phases.

\subsection{Production of $\tilde\chi^0_2 \, \tilde\chi^0_2$ 
           \label{Productionof22}}

\begin{figure}[h]
\setlength{\unitlength}{1cm}
\begin{picture}(10,8)(-0.5,0)
   \put(0,0){\includegraphics{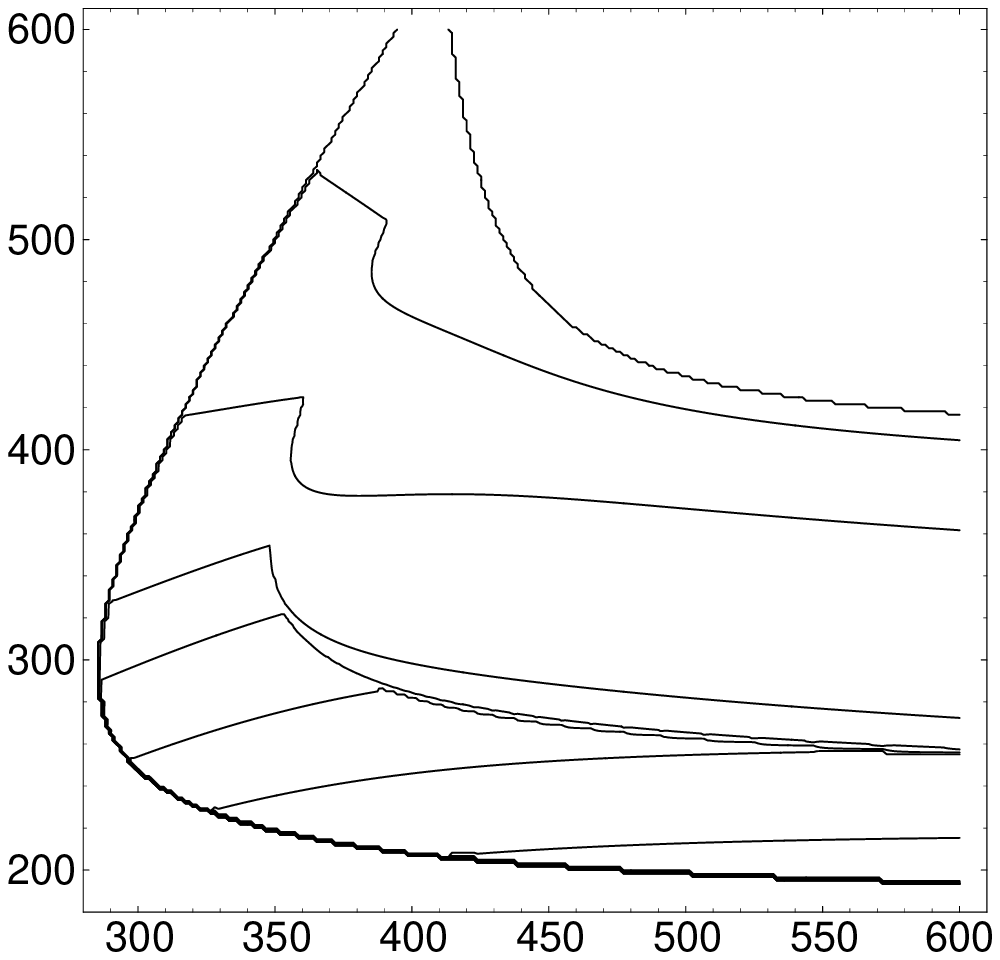}}
	\put(3.5,7.4){\fbox{$\sigma_t$ in fb}}
	\put(5.5,-0.3){$|\mu|$~/GeV}
	\put(0,7.4){$M_2$~/GeV }
	\put(4.5,1.1){\footnotesize 12}
	\put(3.3,1.3){\footnotesize 9}
	\put(2.,1.7){\footnotesize 6}
	\put(1.4,2.2){\footnotesize 3}
	\put(1.15,2.65){\footnotesize 1.5}
	\put(3.8,2.3){\footnotesize 1.5}
	\put(3.5,3.2){\footnotesize 0.3}
	\put(3.,4.2){\footnotesize 0.03}
  	\put(5.5,6){\begin{picture}(1,1)(0,0)
			\CArc(0,0)(7,0,380)
			\Text(0,0)[c]{{\footnotesize A}}
	\end{picture}}
			\put(1.3,5.5){\begin{picture}(1,1)(0,0)
			\CArc(0,0)(7,0,380)
			\Text(0,0)[c]{{\footnotesize B}}
		\end{picture}}
\put(0.5,-.3){Fig.~\ref{plot_22}a}
	\put(8,0){\includegraphics{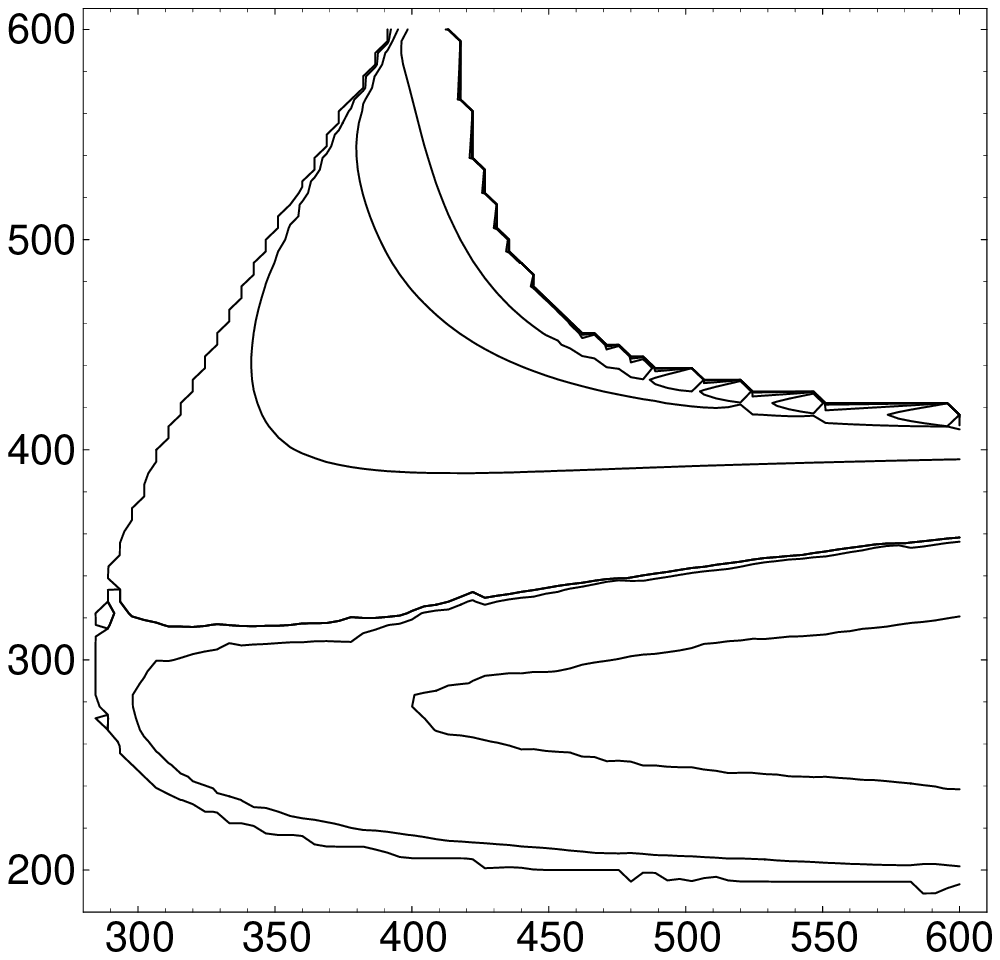}}
	\put(11.,7.4){\fbox{${\mathcal A}_{\ell}$ in \% }}
	\put(13.5,-.3){$|\mu|$~/GeV}
	\put(8,7.4){$M_2$~/GeV }
	\put(11.05,6.0){\footnotesize 3}
	\put(10.75,5.3){\footnotesize 1}
	\put(10.05,3.9){\footnotesize 0.2}
	\put(9.4,2.7){\footnotesize 0}
	\put(9.3,1.9){\footnotesize -0.01}
	\put(11.2,1.9){\footnotesize -0.05}
	  	\put(13.5,6){\begin{picture}(1,1)(0,0)
			\CArc(0,0)(7,0,380)
			\Text(0,0)[c]{{\footnotesize A}}
	\end{picture}}
			\put(9.3,5.5){\begin{picture}(1,1)(0,0)
			\CArc(0,0)(7,0,380)
			\Text(0,0)[c]{{\footnotesize B}}
		\end{picture}}
	\put(8.5,-.3){Fig.~\ref{plot_22}b}
\end{picture}
\vspace*{.5cm}
\caption{
	Contour lines of 
	$\sigma_t=\sigma(e^+e^-\to\tilde\chi^0_2\tilde\chi^0_2 ) 
	\times{\rm BR}(\tilde{\chi}^0_2 \to Z\tilde{\chi}_1^0)\times
	{\rm BR}(Z\to\ell\bar\ell)$ (\ref{plot_22}a),
	and the asymmetry ${\mathcal A}_{\ell}$ (\ref{plot_22}b)
	in the $|\mu|$--$M_2$ plane for $\varphi_{M_1}=0.5\pi $, 
	$\varphi_{\mu}=0$, taking  $\tan \beta=10$, $m_0=300$ GeV,
	$\sqrt{s}=800$ GeV and $(P_{e^-},P_{e^+})=(-0.8,0.6)$.
	The area A (B) is kinematically forbidden by
	$m_{\tilde\chi^0_2}+m_{\tilde\chi^0_2}>\sqrt{s}$
	$(m_{Z}+m_{\tilde\chi^0_1}> m_{\tilde\chi^0_2})$.
	\label{plot_22}}
\end{figure}

In Fig.~\ref{plot_22}a we show the cross section 
$\sigma_t=\sigma(e^+e^-\to\tilde\chi^0_2\tilde\chi^0_2 ) 
\times{\rm BR}(\tilde{\chi}^0_2 \to Z\tilde{\chi}_1^0)\times
{\rm BR}(Z\to\ell\bar\ell)$ in the $|\mu|$--$M_2$ plane
for $\varphi_{\mu}=0$ and $\varphi_{M_1}=0.5\pi$.
The production cross section
$\sigma(e^+e^-\to\tilde\chi^0_2\tilde\chi^0_2 )$, 
which is not shown, is enhanced by the choice
$(P_{e^-},P_{e^+})=(-0.8,0.6)$ and reaches values up to 130 fb.
The branching ratio ${\rm BR}(\tilde{\chi}^0_2 \to
Z\tilde{\chi}_1^0)$, shown in Fig.~\ref{plot_12}b, can be 100\%.
However, due to the small branching ratio  BR($Z\to\ell\bar\ell$) = 0.1,
the cross section shown in Fig.~\ref{plot_22}a does not exceed 13 fb.

If two equal neutralinos are produced,
the  CP sensitive transverse polarization of the neutralinos 
perpendicular to  the production plane vanishes, 
$\Sigma^{2}_P=0 $ in Eq.~(\ref{asym}).
However, the asymmetry ${\mathcal A}_{f}$ need not vanish,
because there are CP sensitive contributions 
from the  neutralino decay process,
terms with $a=1,3$ in Eq.~(\ref{adependence}).
In Fig.~\ref{plot_22}b we show the $|\mu|$ and $M_2$ dependence 
of the asymmetry ${\mathcal A}_{\ell}$, which 
reaches more than  $3 \% $ for $\varphi_{M_1}=0.5\pi $ and 
$\varphi_{\mu}=0$. Along the zero contour in  Fig.~\ref{plot_22}b
the contribution to ${\mathcal A}_{\ell}$ which is proportional
to $ \Sigma_P^1$, see Eq.~\ref{asym}, cancels that 
which is proportional to $ \Sigma_P^3$.
As the largest values of ${\mathcal A}_{\ell}\gsim 0.2\% $ 
and ${\mathcal A}_{q}\gsim 1\% $
lie in a region of the $|\mu|$--$M_2$ plane where $\sigma_t\lsim0.3$ fb,
it will be difficult to measure ${\mathcal A}_{f}$  
in a statistically significant way.
We also studied the $\varphi_{\mu}$ dependence of 
${\mathcal A}_{\ell}$. In the $|\mu|$--$M_2$ plane for $\varphi_{M_1}=0$ 
and $\varphi_{\mu}=0.5\pi$ we found $|{\mathcal A}_{\ell}|<0.5\%$, 
and thus the influence of $\varphi_{\mu}$ is also small.

\begin{figure}[h]
\begin{minipage}{0.5\textwidth}
\begin{picture}(10,7)(0,0)
   \put(-1,8){\includegraphics{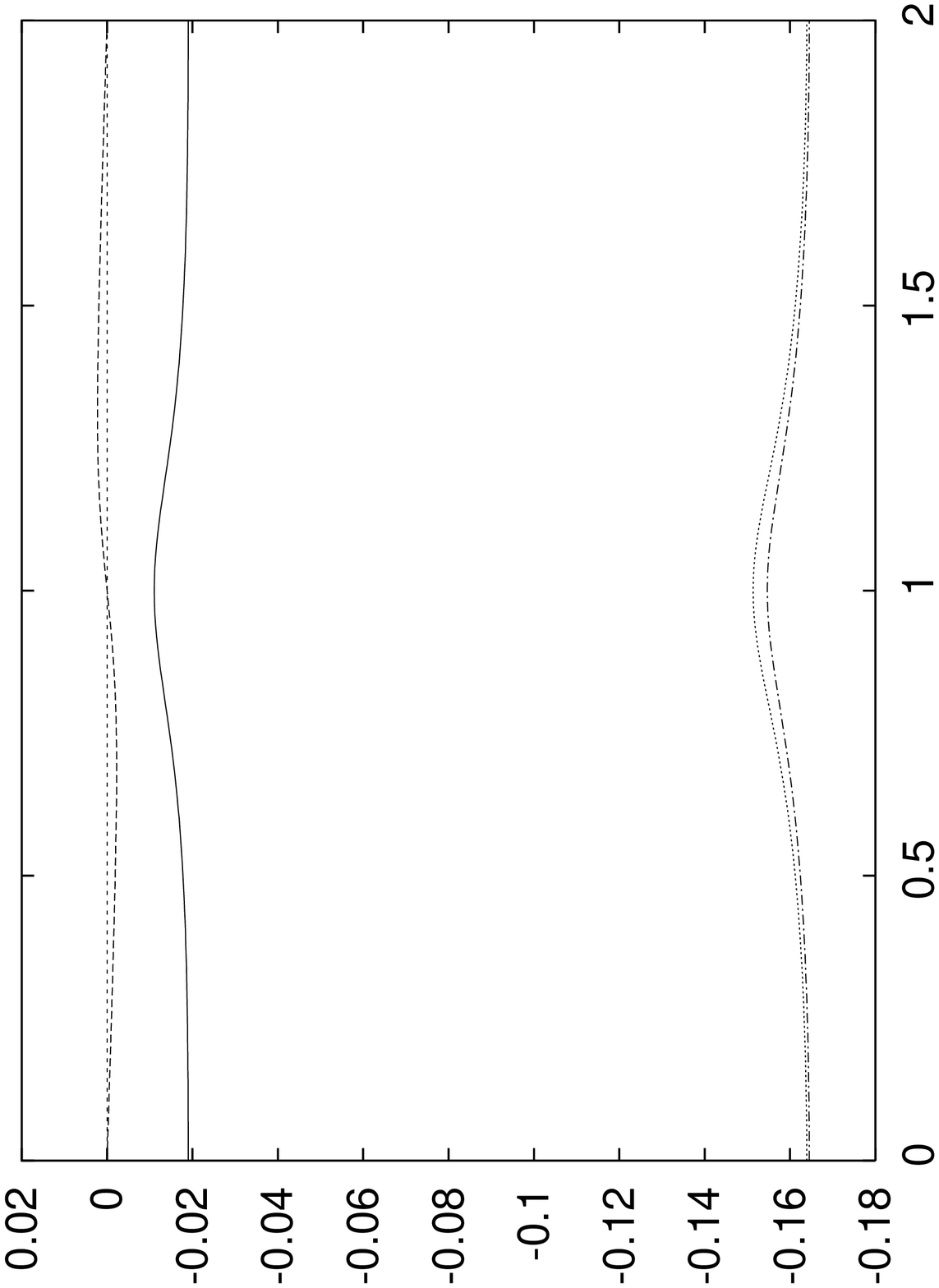}}
	\put(2.5,7.3){\fbox{ matrix elements}}
	\put(6.3,-0.3){$\varphi_{M_1}/\pi $}
	\put(3.8,5.4){\footnotesize $V_1 $}
	\put(5.05,6.){\footnotesize $V_3=0 $}
	\put(5.3,6.5){\footnotesize $V_2$}
	\put(3.8,1.7){\footnotesize $ T_{11}$}
	\put(3.8,1.05){\footnotesize $ T_{22}$}
\end{picture}
\caption{
	Dependence on $\varphi_{M_1}$ of the 
	vector $(V_i)$ and tensor $(T_{ii})$ 
	elements of the $Z$ density matrix  $<\rho(Z)>$,
	for $e^+e^-\to\tilde\chi^0_2\tilde\chi^0_2;
	\tilde{\chi}^0_2 \to Z\tilde{\chi}_1^0$, 
	for $M_2=250$ GeV and $|\mu|=400$ GeV, 
	taking $\varphi_{\mu}=0$, $\tan \beta=10$, $m_0=300$ GeV,
	$\sqrt{s}=800$ GeV and $(P_{e^-},P_{e^+})=(-0.8,0.6)$.
	\label{plot_6}}
\end{minipage}
\end{figure}

In Fig.~\ref{plot_6} we show the 
$\varphi_{M_1}$ dependence of the vector $(V_i)$ and tensor $(T_{ii})$ 
elements of the $Z$ density matrix  $<\rho(Z)>$.
Because there are only CP sensitive contributions from the  neutralino 
decay process, $V_2$ is only zero at $\varphi_{M_1}=0,\pi$ and
$V_1$ has one maximum at $\varphi_{M_1}=\pi$, 
compared to the elements shown in Fig.~\ref{plot_4}.
In addition, in Fig.~\ref{plot_6} the vector elements $V_1$ and 
$V_2$ are much smaller than the 
tensor elements $T_{11}\approx T_{22}$, compared to Fig.~\ref{plot_4}.
The smallness of the vector element $V_2$ accounts for the
smallness of the asymmetry $|{\mathcal A}_{\ell}|<0.05\%$.
Furthermore, the other elements are small, i.e. $T_{13}<10^{-6}$
and $V_3=0$.

%
%
%

\subsection{Production of $\tilde\chi^0_1 \, \tilde\chi^0_3$ }

\begin{figure}[h]
\setlength{\unitlength}{1cm}
\begin{picture}(10,7.8)(-0.5,0)
   \put(0,0){\includegraphics{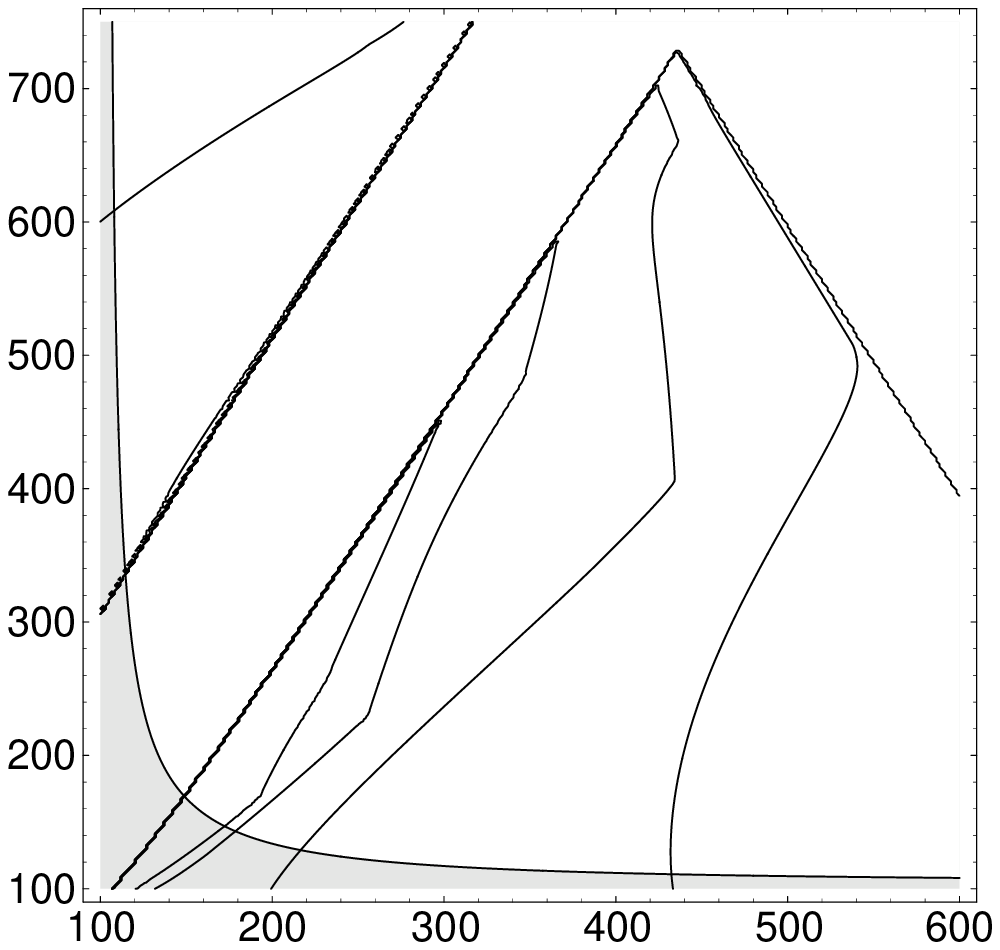}}
	\put(3.5,7.4){\fbox{$\sigma_t$ in fb}}
	\put(5.5,-0.3){$|\mu|$~/GeV}
	\put(0,7.4){$M_2$~/GeV }
	\put(1.35,6.3){\footnotesize 0.03}
	\put(1.,4.){\footnotesize 0.3}
	\put(1.95,1.8){\scriptsize 3}
	\put(3.05,2.8){\footnotesize 1.5}
	\put(4.85,3.8){\footnotesize 0.3}
	\put(5.1,2.0){\footnotesize 0.03}
  	\put(6.2,6){\begin{picture}(1,1)(0,0)
			\CArc(0,0)(7,0,380)
			\Text(0,0)[c]{{\footnotesize A}}
	\end{picture}}
			\put(2.6,4.5){\begin{picture}(1,1)(0,0)
			\CArc(0,0)(7,0,380)
			\Text(0,0)[c]{{\footnotesize B}}
		\end{picture}}
\put(0.5,-.3){Fig.~\ref{plot_13}a}
	\put(8,0){\includegraphics{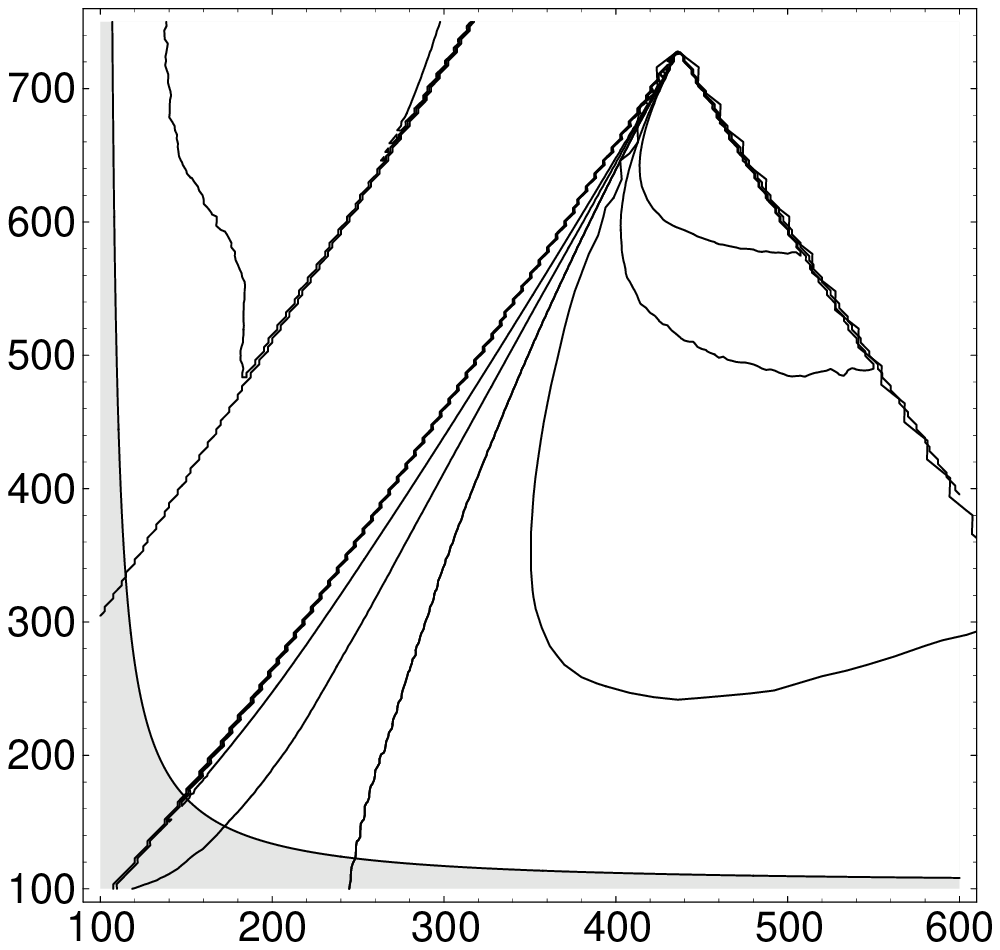}}
	\put(11.,7.4){\fbox{${\mathcal A}_{\ell}$ in \% }}
	\put(13.5,-.3){$|\mu|$~/GeV}
	\put(8,7.4){$M_2$~/GeV }
	\put(10.8,6.55){\footnotesize 1}
	\put(9.4,5.9){\footnotesize 0.5}
	\put(12.75,5.45){\footnotesize -1.3}
	\put(12.9,4.7){\footnotesize -1}
	\put(12.1,2.3){\footnotesize -0.5}
	\put(10.7,1.1){\footnotesize 0}
	\put(10.15,1.6){\footnotesize 0.5}
	\put(10.0,2.){\scriptsize 1}
	  	\put(14.2,6){\begin{picture}(1,1)(0,0)
			\CArc(0,0)(7,0,380)
			\Text(0,0)[c]{{\footnotesize A}}
	\end{picture}}
			\put(10.6,4.5){\begin{picture}(1,1)(0,0)
			\CArc(0,0)(7,0,380)
			\Text(0,0)[c]{{\footnotesize B}}
		\end{picture}}
	\put(8.5,-.3){Fig.~\ref{plot_13}b}
\end{picture}
\caption{
	Contour lines of 
	$\sigma_t=\sigma(e^+e^-\to\tilde\chi^0_1\tilde\chi^0_3) 
	\times{\rm BR}(\tilde{\chi}^0_3 \to Z\tilde{\chi}_1^0)\times
	{\rm BR}(Z\to\ell\bar\ell)$ (\ref{plot_13}a),
	and the asymmetry ${\mathcal A}_{\ell}$ (\ref{plot_13}b)
	in the $|\mu|$--$M_2$ plane for $\varphi_{M_1}=0.5\pi $, 
	$\varphi_{\mu}=0$, taking  $\tan \beta=10$, $m_0=300$ GeV,
	$\sqrt{s}=800$ GeV and $(P_{e^-},P_{e^+})=(0.8,-0.6)$.
	The area A (B) is kinematically forbidden by
	$m_{\tilde\chi^0_1}+m_{\tilde\chi^0_3}>\sqrt{s}$
	$(m_{Z}+m_{\tilde\chi^0_1}> m_{\tilde\chi^0_3})$.
		The gray area is excluded by $m_{\tilde\chi_1^{\pm}}<104$ GeV.
	\label{plot_13}}
\end{figure}

In Fig.~\ref{plot_13}a we show the cross section 
$\sigma_t=\sigma(e^+e^-\to\tilde\chi^0_1\tilde\chi^0_3) 
\times{\rm BR}(\tilde{\chi}^0_3 \to Z\tilde{\chi}_1^0)\times
{\rm BR}(Z\to\ell\bar\ell)$ in the $|\mu|$--$M_2$ plane
for $\varphi_{\mu}=0$ and $\varphi_{M_1}=0.5\pi$.
The production cross section 
$\sigma(e^+e^-\to\tilde\chi^0_1\tilde\chi^0_3)$,
which is not shown, is enhanced by the choice 
$(P_{e^-},P_{e^+})=(0.8,-0.6)$ and reaches up to 50 fb.
The branching ratio ${\rm BR}(\tilde{\chi}^0_3 \to
Z\tilde{\chi}_1^0)$, which is not shown, can be 1.
However, due to the small branching ratio  BR($Z\to\ell\bar\ell$) = 0.1,
the cross section shown in Fig.~\ref{plot_13}a does not exceed 5 fb.
In Fig.~\ref{plot_13}b we show the $|\mu|$--$M_2$ dependence 
of the asymmetry ${\mathcal A}_{\ell}$.  
The asymmetry $|{\mathcal A}_{\ell}|$ reaches  $1.3 \% $
at its maximum, however in a region, where $\sigma_t<0.3$ fb,
the asymmetry ${\mathcal A}_{\ell}$ thus cannot be measured. 
We also studied the $\varphi_{\mu}$ dependence of 
${\mathcal A}_{\ell}$. In the $|\mu|$--$M_2$ plane for $\varphi_{M_1}=0$ 
and $\varphi_{\mu}=0.5\pi$ we found $|{\mathcal A}_{\ell}|<0.7\%$. 
%

\subsection{Production of $\tilde\chi^0_2 \, \tilde\chi^0_3$ }

\begin{figure}[h]
\setlength{\unitlength}{1cm}
\begin{picture}(10,7.6)(-0.5,0)
   \put(0,0){\includegraphics{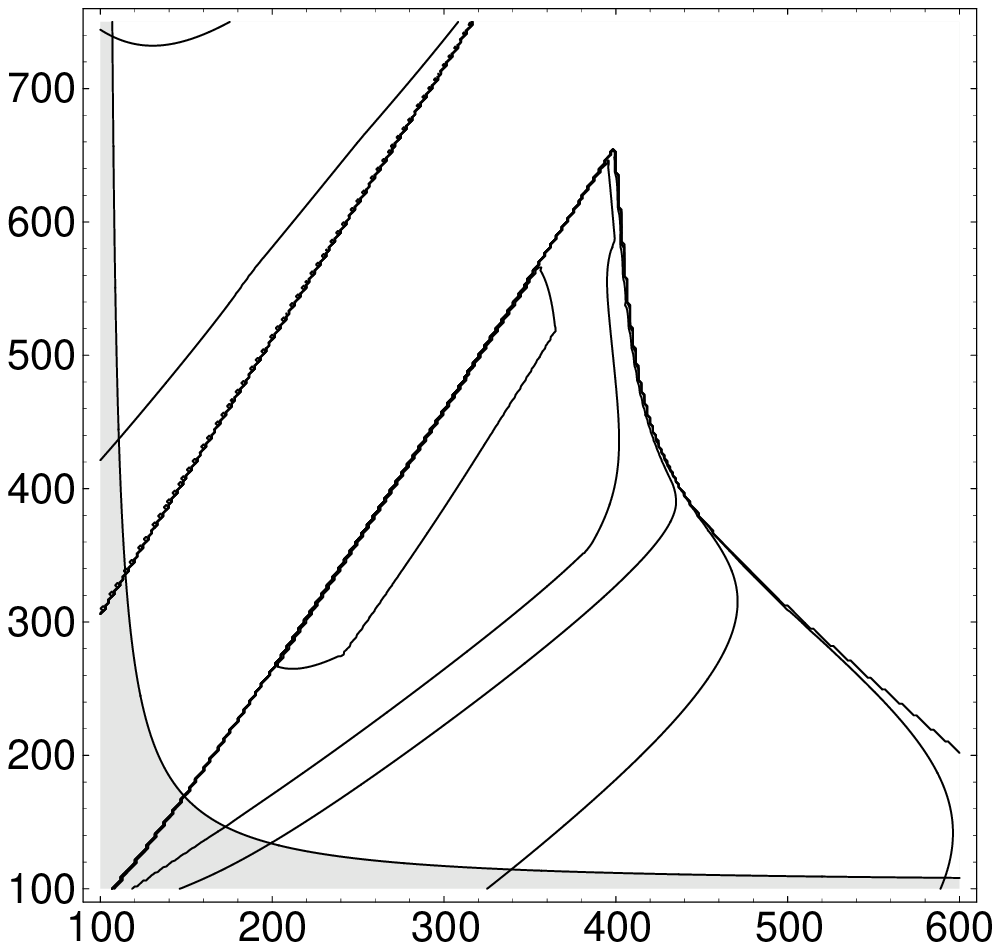}}
	\put(3.5,7.4){\fbox{$\sigma_t$ in fb}}
	\put(5.5,-0.3){$|\mu|$~/GeV}
	\put(0,7.4){$M_2$~/GeV }
	\put(1.4,6.5){\footnotesize 0.03}
	\put(1.4,5.4){\footnotesize 0.3}
	\put(2.8,3.4){\footnotesize 4.5}
	\put(3.8,3.4){\footnotesize 1.5}
	\put(4.25,2.4){\footnotesize 0.3}
	\put(5.1,2.0){\footnotesize 0.03}
	\put(5.9,1.0){\footnotesize 0.003}
  	\put(5.52,5){\begin{picture}(1,1)(0,0)
			\CArc(0,0)(7,0,380)
			\Text(0,0)[c]{{\footnotesize A}}
	\end{picture}}
			\put(2.7,4.5){\begin{picture}(1,1)(0,0)
			\CArc(0,0)(7,0,380)
			\Text(0,0)[c]{{\footnotesize B}}
		\end{picture}}
\put(0.5,-.3){Fig.~\ref{plot_23}a}
	\put(8,0){\includegraphics{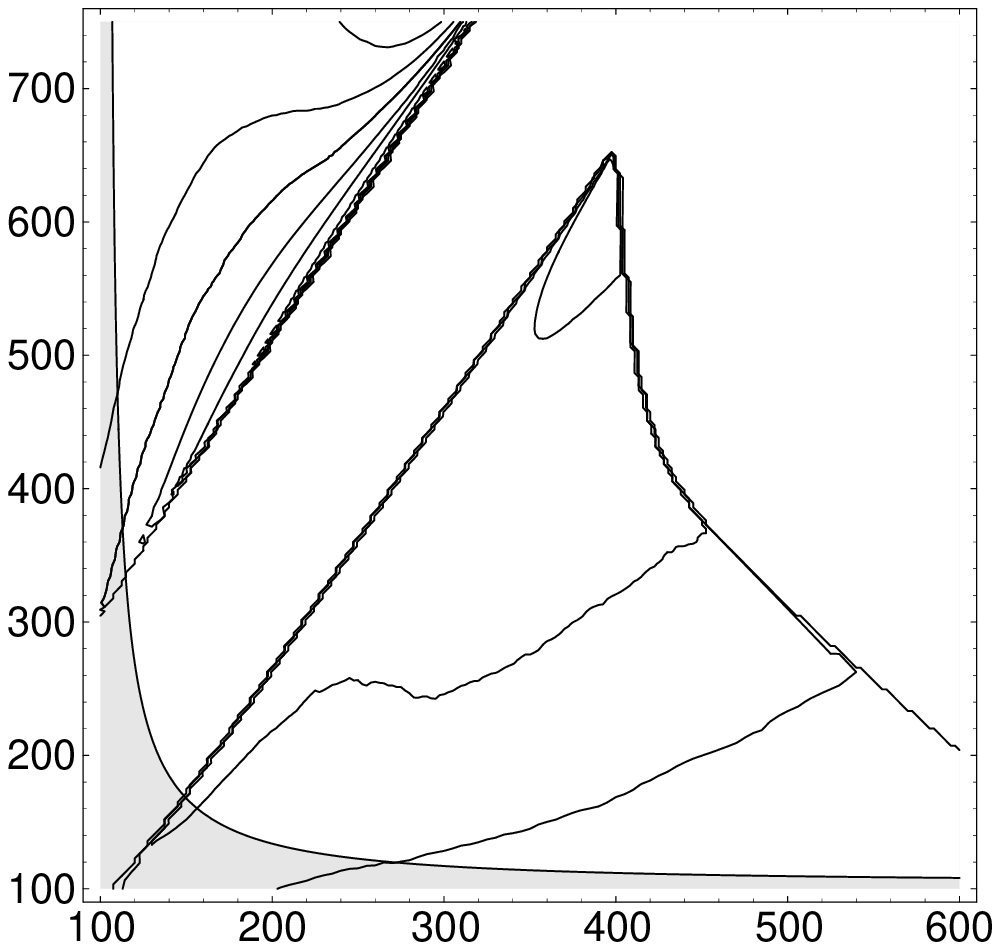}}
	\put(11.,7.4){\fbox{${\mathcal A}_{\ell}$ in \% }}
	\put(13.5,-.3){$|\mu|$~/GeV}
	\put(8,7.4){$M_2$~/GeV }
	\put(10.7,6.75){\scriptsize 1}
	\put(9.3,6.15){\footnotesize 0.5}
	\put(9.58,5.4){\footnotesize 0}
	\put(9.85,4.85){\scriptsize -1}
	\put(9.73,5.15){\scriptsize -.5}
	\put(12.,4.9){\footnotesize -2}
	\put(12.15,2.9){\footnotesize -1}
	\put(13.,1.95){\footnotesize -0.5}
	  	\put(13.8,5){\begin{picture}(1,1)(0,0)
			\CArc(0,0)(7,0,380)
			\Text(0,0)[c]{{\footnotesize A}}
	\end{picture}}
			\put(10.7,4.5){\begin{picture}(1,1)(0,0)
			\CArc(0,0)(7,0,380)
			\Text(0,0)[c]{{\footnotesize B}}
		\end{picture}}
	\put(8.5,-.3){Fig.~\ref{plot_23}b}
\end{picture}
\caption{
	Contour lines of 
	$\sigma_t=\sigma(e^+e^-\to\tilde\chi^0_2\tilde\chi^0_3) 
	\times{\rm BR}(\tilde{\chi}^0_3 \to Z\tilde{\chi}_1^0)\times
	{\rm BR}(Z\to\ell\bar\ell)$ (\ref{plot_23}a),
	and the asymmetry ${\mathcal A}_{\ell}$ (\ref{plot_23}b)
	in the $|\mu|$--$M_2$ plane for $\varphi_{M_1}=0.5\pi $, 
	$\varphi_{\mu}=0$, taking  $\tan \beta=10$, $m_0=300$ GeV,
	$\sqrt{s}=800$ GeV and $(P_{e^-},P_{e^+})=(0.8,-0.6)$.
	The area A (B) is kinematically forbidden by
	$m_{\tilde\chi^0_2}+m_{\tilde\chi^0_3}>\sqrt{s}$
	$(m_{Z}+m_{\tilde\chi^0_1}> m_{\tilde\chi^0_3})$.
		The gray area is excluded by $m_{\tilde\chi_1^{\pm}}<104$ GeV.
	\label{plot_23}}
\end{figure}

For the  process 
$e^+e^-\to\tilde\chi^0_2\tilde\chi^0_3$
we discuss the decay  $\tilde\chi^0_3\to Z\tilde{\chi}_1^0$ of the
heavier neutralino which has a larger kinematically allowed region
than that of $\tilde\chi^0_2\to Z\tilde{\chi}_1^0$. Similar to 
$\tilde\chi^0_1 \, \tilde\chi^0_3$ production and decay, the cross
section $\sigma(e^+e^-\to\tilde\chi^0_2\tilde\chi^0_3)$ reaches
values up to 50 fb for a beam polarization of
$(P_{e^-},P_{e^+})=(0.8,-0.6)$. The cross section for the complete
process $\sigma_t=\sigma(e^+e^-\to\tilde\chi^0_2\tilde\chi^0_3) 
\times{\rm BR}(\tilde{\chi}^0_3 \to Z\tilde{\chi}_1^0)\times
{\rm BR}(Z\to\ell\bar\ell)$ attains values up to 5 fb in the 
$|\mu|$--$M_2$ plane, see Fig.~\ref{plot_23}a.

The asymmetry ${\mathcal A}_{\ell}$, Fig.~\ref{plot_23}b,
is somewhat larger than the asymmetry for  
$\tilde\chi^0_1 \, \tilde\chi^0_3$ production and decay, and
reaches at its maximum 2\%. Although in the respective region the cross
section is also a bit larger, $\sigma_t \lsim 4$ fb, 
it will be difficult to measure ${\mathcal A}_{\ell}$.
For example taking $|\mu|=380$ GeV, $M_2=560$ GeV and
$(\varphi_{M_1},\varphi_{\mu})=(0.5 \pi,0)$,
we have $S_{\ell}\approx1$, for ${\mathcal L}=500~{\rm fb}^{-1}$.
However for the hadronic decays of the $Z$ we have $S_{b(c)}\approx 8(5)$
and thus ${\mathcal A}_{b(c)}$ could be measured
for $\tilde\chi^0_1 \, \tilde\chi^0_3$ production.
Concerning the $\varphi_{\mu}$ dependence of 
${\mathcal A}_{\ell}$ we found that $|{\mathcal A}_{\ell}|\lsim1\%$
in regions of the $|\mu|$--$M_2$ plane where $\sigma_t \lsim 0.5$ fb,
and $|{\mathcal A}_{\ell}|\lsim0.4\%$
in regions where $\sigma_t \lsim 5$ fb, for example 
for $\varphi_{\mu}=0.5\pi$ and $\varphi_{M_1}=0$.
%
%

\section{Summary and conclusions
	\label{Summary and conclusion}}

We have proposed and analyzed CP sensitive observables in 
neutralino production $e^+e^- \to\tilde{\chi}^0_i  \tilde{\chi}^0_j$
and the subsequent two-body decay  of one  neutralino
into the $Z$ boson $\tilde\chi^0_i \to \chi^0_n Z$,
followed by the decay $Z \to \ell \bar\ell $ for $ \ell= e,\mu,\tau$, 
or $Z \to q\bar q$ for $q=c,b$.
The CP sensitive observables are defined by the vector component $V_2$ 
of the $Z$ boson density matrix and the CP asymmetry 
$ {\mathcal A}_{\ell(q)}$, which involves the triple product
${\mathcal T}_{\ell(q)}= \vec p_{e^-}\cdot(\vec p_{\ell(q)} 
\times \vec p_{\bar\ell(\bar q)}).$
The tree level contributions to these observables
are due to correlations of the neutralino $\tilde\chi^0_i$ spin and 
the $Z$ boson spin. 
In a numerical study of the MSSM parameter space with 
complex $M_1$ and $\mu$ for  
$\tilde{\chi}^0_1  \tilde{\chi}^0_2$,
$\tilde{\chi}^0_2  \tilde{\chi}^0_2$,
$\tilde{\chi}^0_1  \tilde{\chi}^0_3$ and
$\tilde{\chi}^0_2  \tilde{\chi}^0_3$ production,
we have shown that the asymmetry ${\mathcal A}_{\ell}$ 
can go up to 3\%. For the hadronic decays of the Z boson,  
larger asymmetries  are obtained with
${\mathcal A}_{c(b)} \simeq6.3(4.5)\times {\mathcal A}_{\ell}$.
By analyzing their statistical errors, we found that the
asymmetries ${\mathcal A}_{c(b)}$
could be accessible in future electron positron linear collider 
experiments in the 500-800 GeV range with high luminosity and 
longitudinally polarized beams.

\section{Acknowledgments}

We thank S. Hesselbach and T. Kernreiter for useful discussions.
This work was supported by the `Fonds zur
F\"orderung der wissenschaftlichen Forschung' (FWF) of Austria, projects
No. P13139-PHY and No. P16592-N02, by the European Community's
Human Potential Programme under contract HPRN-CT-2000-00149
and HPRN-CT-2000-00148 and by  Spanish grants BFM2002-00345.
This work was also supported by the 'Deutsche Forschungsgemeinschaft'
(DFG) under contract Fr 1064/5-1.
OK was supported  by the EU Research Training Site 
contract HPMT-2000-00124.

\newpage

\begin{appendix}
	\noindent{\Large\bf Appendix}

\setcounter{equation}{0}
\renewcommand{\thesubsection}{\Alph{section}.\arabic{subsection}}
\renewcommand{\theequation}{\Alph{section}.\arabic{equation}}
\section{Coordinate frame and spin vectors
     \label{Representation of momentum and spin vectors}}
\setcounter{equation}{0}

We choose a coordinate frame in the laboratory system
such that the momentum of neutralino
$\tilde \chi ^0_j$ points in the $z$-direction
(in our definitions we follow closely \cite{gudi1}). 
The scattering  angle is 
$\theta \angle (\vec p_{e^-},\vec p_{\chi_j})$ and 
the azimuth $\phi$ can be chosen zero. The momenta are given by: 
   \begin{eqnarray}
  && p_{e^-} = E_b(1,-\sin\theta,0, \cos\theta),\quad
     p_{e^+} = E_b(1, \sin\theta,0,-\cos\theta),\\
 &&  p_{\chi_i} = (E_{\chi_i},0,0,-q),\quad
     p_{\chi_j} = (E_{\chi_j},0,0, q),
   \end{eqnarray}
with the beam energy $E_b=\sqrt{s}/2$ and
\begin{eqnarray}
 &&   E_{\chi_i} =\frac{s+m_{\chi_i}^2-m_{\chi_j}^2}{2 \sqrt{s}},\quad
    E_{\chi_j} =\frac{s+m_{\chi_j}^2-m_{\chi_i}^2}{2 \sqrt{s}},\quad
      q =\frac{\lambda^{\frac{1}{2}}
             (s,m_{\chi_i}^2,m_{\chi_j}^2)}{2 \sqrt{s}}, 
\end{eqnarray}
where $m_{\chi_i}, m_{\chi_j}$ are the masses of the neutralinos and 
$\lambda(x,y,z) = x^2+y^2+z^2-2(xy+xz+yz)$.
We choose the  three spin  vectors $s^{a,\mu}_{\chi_i}$ ($a=1,2,3$) 
of the neutralino in the laboratory system by:
\begin{eqnarray}
	&&  s^1_{\chi_i}=(0,-1,0,0),\quad
    s^2_{\chi_i}=(0,0,1,0),\quad
    s^3_{\chi_i}=\frac{1}{m_{\chi_i}}(q,0,0,-E_{\chi_i}).
\end{eqnarray} 
Together with  
$p_{\chi_i}^{\mu}/m_{\chi_i}$ they form an orthonormal set.
For the two-body decay $\tilde\chi_i^0\to\tilde\chi_n^0Z$
the decay angle 
$\theta_{1} \angle (\vec p_{\chi_i},\vec p_{Z})$
is constrained  by $\sin\theta^{\rm max}_{1}= q^0/q$
for $q>q^0$,
where $q^0=\lambda^{\frac{1}{2}}(m^2_{\chi_i},m^2_Z,m^2_{\chi_n})/2m_Z$
is the neutralino momentum if the $Z$ boson is produced at rest.
In this case there are two solutions 
\begin{eqnarray}
| \vec p^{\pm}_Z|= \frac{
(m^2_{\chi_i}+m^2_Z-m^2_{\chi_n})q\cos\theta_{1}\pm
E_{\chi_i}\sqrt{\lambda(m^2_{\chi_i},m^2_Z,m^2_{\chi_n})-
	 4q^2~m^2_Z~(1-\cos^2\theta_{1})}}
	{2q^2 (1-\cos^2\theta_{1})+2 m^2_{\chi_i}}.
\end{eqnarray}
If $q^0>q$, $\theta_{1}$ is not 
constrained and there is only the physical solution 
$ |\vec p^+_Z|$ left. The momenta in the laboratory system are
 \begin{eqnarray}
&&   p_{Z}^{\pm} = (                        E_{Z}^{\pm},
            -|\vec p_{Z}^{\pm}| \sin \theta_{1} \cos \phi_{1},
             |\vec p_{Z}^{\pm}| \sin \theta_{1} \sin \phi_{1},
				 -|\vec p_{Z}^{\pm}| \cos \theta_{1}), \\
&& p_{\bar f} = (                        E_{\bar f},
            -|\vec p_{\bar f}| \sin \theta_{2} \cos \phi_{2},
             |\vec p_{\bar f}| \sin \theta_{2} \sin \phi_{2},
				-|\vec p_{\bar f}| \cos \theta_{2}),\\
&& E_{\bar f} = |\vec p_{\bar f}| = 
\frac{m_Z^2}{2(E_{Z}^{\pm}-|\vec p_{Z}^{\pm}|\cos\theta_{D_2})},
 \end{eqnarray}
with $\theta_{2} \angle (\vec p_{\chi_i},\vec p_{\bar f})$
and the decay angle 
$\theta_{D_2} \angle (\vec p_{Z},\vec p_{\bar f})$ given by:
 \begin{equation}
\cos\theta_{D_2}=\cos\theta_{1}\cos\theta_{2}+
\sin\theta_{1}\sin\theta_{2}\cos(\phi_{2}-\phi_{1}). 
  \end{equation}
The spin vectors $t^{c,\mu}_Z$ ($c=1,2,3$) of the $Z$  boson
in the laboratory system are chosen by
\begin{equation}\label{defoft}
t^1_Z=\left(0,\frac{{\vec t}^2_Z
\times{\vec t}_Z^3}{|{\vec t}_Z^2\times{\vec t}^3_Z|}\right),\quad
t^2_Z=\left(0,
\frac{{\vec p}_{e^-}\times{\vec p}_Z}{|{\vec p}_{e^-}
\times{\vec p}_Z|}\right),\quad
t^3_Z=\frac{1}{m_Z}
\left(|{\vec p}_Z|, E_Z \frac{ {\vec p}_Z}{|{\vec p}_Z|} \right).
  \end{equation}
The spin vectors and $p_Z^{\mu}/m_Z$ form an orthonormal set.
The polarization vectors 
$\varepsilon^{\lambda_k,\mu}$ 
for helicities $\lambda_k=-1,0,+1$ of the $Z$ boson
are defined by:
\begin{equation}\label{circularbasis}
	\varepsilon^-={\textstyle \frac{1}{\sqrt 2}}(t^1_Z-i t^2_Z);
	\quad \varepsilon^0=t^3_Z; \quad
	\varepsilon^+=-{\textstyle \frac{1}{\sqrt 2}}(t^1_Z+i t^2_Z).
\end{equation}

\section{Phase space
     \label{Phase space}}
\setcounter{equation}{0}

The Lorentz invariant phase space element for the neutralino 
production (\ref{production}) and the decay 
chain (\ref{decay_1})-(\ref{decay_2}) can be decomposed
into the two-body  phase space elements:
\begin{eqnarray}
 &&d{\rm Lips}(s,p_{\chi_j },p_{\chi_n},p_{f},p_{\bar f}) =
	 \nonumber \\ 
&&\frac{1}{(2\pi)^2}~d{\rm Lips}(s,p_{\chi_i},p_{\chi_j} )
~d s_{\chi_i} ~\sum_{\pm}d{\rm Lips}(s_{\chi_i},p_{\chi_n},p_{Z}^{\pm})
 ~d s_{Z}~d{\rm Lips}(s_{Z},p_{f},p_{\bar f}),\label{Lips}
 \end{eqnarray}
\begin{eqnarray}
	d{\rm Lips}(s,p_{\chi_i },p_{\chi_j })&=&
	\frac{q}{8\pi\sqrt{s}}~\sin\theta~ d\theta, \\
	d{\rm Lips}(s_{\chi_i},p_{\chi_n},p_Z^{\pm})&=&
\frac{1}{2(2\pi)^2}~
\frac{|\vec p_Z^{\pm}|^2}{2|E_Z^{\pm}~q\cos\theta_1-
	E_{\chi_i}~|\vec p^{\pm}_Z||}~d\Omega_1,\\
	d{\rm Lips}(s_{Z},p_{f},p_{\bar f})&=&
\frac{1}{2(2\pi)^2}~\frac{|\vec p_{\bar f}|^2}{m_Z^2}
	~d\Omega_2,
\end{eqnarray}
with $s_{\chi_i}=p^2_{\chi_i}$, $s_{Z}=p^2_{Z}$ and 
$ d\Omega_i=\sin\theta_i~ d\theta_i~ d\phi_i$.
We use the narrow width approximation for the propagators:
$\int|\Delta(\tilde\chi^0_i)|^2 $ $ d s_{\chi_i} = 
\frac{\pi}{m_{\chi_i}\Gamma_{\chi_i}}, ~
\int|\Delta(Z)|^2 d s_{Z} = 
\frac{\pi}{m_{Z}\Gamma_{Z}}$.
The approximation is justified for
$(\Gamma_{\chi_i}/m_{\chi_i})^2\ll1$,
which holds in our case with 
$\Gamma_{\chi_i}\lsim {\mathcal O}(1 {\rm GeV}) $.

\section{Spin matrices
     \label{matrices}}
\setcounter{equation}{0}
In the basis (\ref{circularbasis}) the spin matrices 
$J^{c}$ and the tensor components $J^{cd}$ are 
 \begin{eqnarray}
  J^1=
  \left(
        \begin{array}{rrr}
			  0&\frac{1}{\sqrt{2}}&0\\
			  \frac{1}{\sqrt{2}}&0&\frac{1}{\sqrt{2}}\\
			  0&\frac{1}{\sqrt{2}}&0
        \end{array}
	  \right),
&
  J^2=
  \left(
        \begin{array}{rrr}
			  0&\frac{i}{\sqrt{2}}&0\\
			  -\frac{i}{\sqrt{2}}&0&\frac{i}{\sqrt{2}}
			  \\0&-\frac{i}{\sqrt{2}}&0
        \end{array}
	  \right),
&
  J^3=
  \left(
        \begin{array}{rrr}
         -1&0&0\\0&0&0\\0&0&1
        \end{array}
	  \right), \label{Jcircular1}\\
J^{11}=
  \left(
        \begin{array}{rrr}
			  -\frac{1}{3}&0&1\\
			  0&\frac{2}{3}&0\\
			  1&0&-\frac{1}{3}
        \end{array}
	  \right),
&
J^{22}=
\left(
	\begin{array}{rrr}
	  -\frac{1}{3}&0&-1\\
	  0&\frac{2}{3}&0\\
	  -1&0&-\frac{1}{3}
	  \end{array}
	  \right),
	  &
J^{33}=  
\left(
	\begin{array}{rrr}
	  \frac{2}{3}&0&0\\
	  0&-\frac{4}{3}&0\\
	  0&0&\frac{2}{3}
	  \end{array}
  \right),\label{Jcircular2}\\
J^{12}=
  \left(
        \begin{array}{rrr}
         0&0&i\\0&0&0\\-i&0&0
        \end{array}
	  \right),
&
J^{23}=
  \left(
        \begin{array}{rrr}
			  0&-\frac{i}{\sqrt{2}}&0\\
			  \frac{i}{\sqrt{2}}&0&\frac{i}{\sqrt{2}}
			  \\0&-\frac{i}{\sqrt{2}}&0
        \end{array}
	  \right),
&
J^{13}=
  \left(
        \begin{array}{rrr}
			  0&-\frac{1}{\sqrt{2}}&0\\
			  -\frac{1}{\sqrt{2}}&0&\frac{1}{\sqrt{2}}\\
			  0&\frac{1}{\sqrt{2}}&0
        \end{array}
	  \right). \label{Jcircular3}
\end{eqnarray}

\end{appendix}

\end{document}